\documentclass[twocolumn,twocolappendix]{openjournal}

\usepackage{lipsum}

\usepackage{xcolor}
\usepackage{textgreek}
\usepackage[utf8]{inputenc}
\usepackage[english]{babel}

\usepackage{hyperref}
\hypersetup{
    unicode, 
    colorlinks=true,
    linkcolor=linkcolor,
    citecolor=linkcolor,
    filecolor=linkcolor,
    urlcolor=linkcolor,
}
\usepackage{color,colortbl}
\definecolor{linkcolor}{rgb}{0.0,0.3,0.5}
\usepackage{tensind}
\tensordelimiter{?}
\DeclareGraphicsExtensions{.bmp,.png,.jpg,.pdf}
\usepackage{verbatim}
\usepackage[normalem]{ulem}
\usepackage{orcidlink}
\usepackage{soul}
\usepackage{enumitem}

\graphicspath{ {./figs/} }

\usepackage[T1]{fontenc}

\DeclareRobustCommand{\VAN}[3]{#2}
\let\VANthebibliography\thebibliography
\def\thebibliography{\DeclareRobustCommand{\VAN}[3]{##3}\VANthebibliography}

\usepackage{graphicx}	
\usepackage{adjustbox}  
\usepackage{amsmath}	
\usepackage{wasysym}    
\usepackage{bm}         
\usepackage{xspace}
\usepackage{soul}
\usepackage{multirow}
\makeatletter 
  \patchcmd{\NAT@citex}
    {\@citea\NAT@hyper@{%
      \NAT@nmfmt{\NAT@nm}%
      \hyper@natlinkbreak{\NAT@aysep\NAT@spacechar}{\@citeb\@extra@b@citeb}%
      \NAT@date}}
    {\@citea\NAT@nmfmt{\NAT@nm}%
    \NAT@aysep\NAT@spacechar\NAT@hyper@{\NAT@date}}{}{}

  \patchcmd{\NAT@citex}
    {\@citea\NAT@hyper@{%
      \NAT@nmfmt{\NAT@nm}%
      \hyper@natlinkbreak{\NAT@spacechar\NAT@@open\if*#1*\else#1\NAT@spacechar\fi}%
        {\@citeb\@extra@b@citeb}%
      \NAT@date}}
    {\@citea\NAT@nmfmt{\NAT@nm}%
    \NAT@spacechar\NAT@@open\if*#1*\else#1\NAT@spacechar\fi\NAT@hyper@{\NAT@date}}
    {}{}
\makeatother \usepackage{amssymb}	

\newcommand\Msun{\text{M}_{\astrosun}} 
\newcommand\Zsun{\text{Z}_{\astrosun}} 
\newcommand\HI{H\,\textsc{\lowercase{I}}\xspace} 
\newcommand\HII{H\,\textsc{\lowercase{II}}\xspace} 
\newcommand\thesan{\mbox{\textsc{thesan}}\xspace}
\newcommand\thesanone{\mbox{\textsc{thesan-1}}\xspace}
\newcommand\thesantwo{\mbox{\textsc{thesan-2}}\xspace}
\newcommand\thesanwc{\mbox{\textsc{thesan-wc-2}}\xspace}
\newcommand\thesanhigh{\mbox{\textsc{thesan-high-2}}\xspace}
\newcommand\thesanlow{\mbox{\textsc{thesan-low-2}}\xspace}
\newcommand\thesansdao{\mbox{\textsc{thesan-sdao-2}}\xspace}

\newcommand\arepo{\mbox{\textsc{arepo}}\xspace}
\newcommand\areport{\mbox{\textsc{arepo-rt}}\xspace}
\newcommand\colt{\mbox{\textsc{colt}}\xspace}

\newcommand\fT{$f_{\rm esc} \times \mathcal{T}_{\rm IGM}$\xspace}
\newcommand\LLya{$L_{\rm Ly \alpha}$\xspace}
\newcommand\ergs{$\rm erg\, s^{-1}$\xspace}
\newcommand\orcid[1]{\href{http://orcid.org/#1}{\adjustbox{trim={-.15\width} {0\height} {-.15\width} {0\height},clip}{\includegraphics[height=10pt]{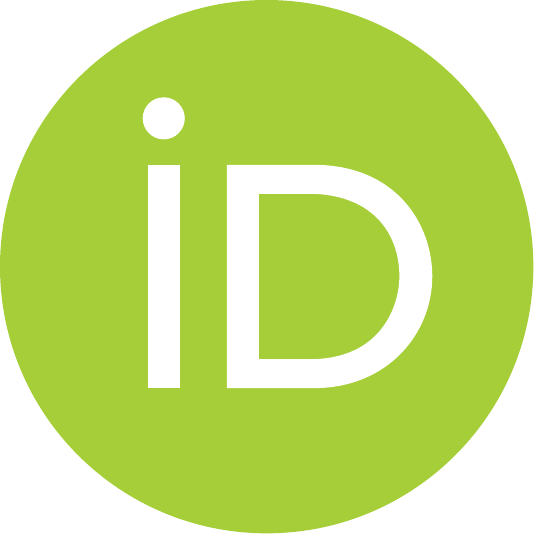}}}}

\begin{document}
\title{The thesan project: Lyman-$\alpha$ emitters as probes of ionized bubble sizes}

\author{Meredith Neyer\orcid{0000-0002-9205-9717}$^{1\,\star}$}
\email[$^\star$ E-mail: ]{mneyer@mit.edu}
\author{Aaron~Smith\orcid{0000-0002-2838-9033}$^{2}$}
\author{Mark~Vogelsberger\orcid{0000-0001-8593-7692}$^{1,3}$}
\author{Luz~\'Angela Garc\'ia\orcid{0000-0003-1235-794X}$^{4}$}
\author{Rahul~Kannan\orcid{0000-0001-6092-2187}$^{5}$}
\author{Enrico~Garaldi\orcid{0000-0002-6021-7020}$^{6}$}
\author{Laura~Keating\orcid{0000-0001-5211-1958}$^{7}$}
\affiliation{$^{1}$Department of Physics $\&$ Kavli Institute for Astrophysics and Space Research, Massachusetts Institute of Technology, Cambridge, MA 02139, USA}
\affiliation{$^{2}$Department of Physics, The University of Texas at Dallas, Richardson, Texas 75080, USA}
\affiliation{$^{3}$The NSF AI Institute for Artificial Intelligence and Fundamental Interactions, Massachusetts Institute of Technology, Cambridge, MA 02139, USA}
\affiliation{$^{4}$Universidad ECCI, Cra. 19 No. 49-20, Bogot\'a, Colombia, C\'odigo Postal 111311}
\affiliation{$^{5}$Department of Physics and Astronomy, York University, 4700 Keele Street, Toronto, ON M3J 1P3, Canada}
\affiliation{$^{6}$Kavli IPMU (WPI), UTIAS, The University of Tokyo, Kashiwa, Chiba 277-8583, Japan}
\affiliation{$^{7}$Institute for Astronomy, University of Edinburgh, Blackford Hill, Edinburgh, EH9 3HJ, UK}




\begin{abstract}
  We use the \textsc{thesan} radiation-hydrodynamics simulations to investigate how Lyman-$\alpha$ emitters (LAEs) trace ionized bubble sizes during the Epoch of Reionization. We generate realistic LAE catalogs by combining accurate intrinsic Ly$\alpha$ production and intergalactic transmission with an empirical model for dust absorption and gas outflows. By calibrating to observationally-constrained Ly$\alpha$ luminosity functions, we reproduce the rapid decline in Ly$\alpha$ visibility toward higher redshifts while revealing mild tensions in LAE fractions near the end of reionization. Before the midpoint of reionization, galaxies within larger line-of-sight bubbles ($\gtrsim 10$\,cMpc) have higher observed Ly$\alpha$ luminosity and equivalent width (EW), demonstrating that the evolving LAE fraction provides a practical statistical tracer for bubble size. These correlations weaken as percolation progresses and the IGM becomes increasingly ionized.
  In LAE selected samples with $L_{\text{Ly}\alpha} > 10^{41.5}\,\text{erg\,s}^{-1}$,
  Ly$\alpha$ properties correlate with bubble size more strongly than UV magnitude, especially at $z \gtrsim 7$. This simulation-based framework maps LAE selections to bubble-size statistics, clarifies biases in more idealized models, and will supply public catalogs to interpret current and forthcoming \textit{JWST} and narrow-band LAE surveys in terms of the evolving topology of reionization.
\end{abstract}


\maketitle



\section{Introduction}

Within the first billion years after the Big Bang, the Universe underwent a pivotal transformation known as the Epoch of Reionization (EoR). During this period, high-energy photons emitted by the first generations of galaxies ionized the surrounding neutral hydrogen gas in the intergalactic medium (IGM), gradually ending the cosmic dark ages \citep{Shapiro1987, BarkanaLoeb2001, Furlanetto2006b, Wise2019}. The EoR occurred at redshifts of approximately $z \approx 5 - 20$, which can be characterized by the formation and growth of ionized hydrogen bubbles around early luminous sources. These \HII regions expanded and eventually merged and percolated, leading to a fully ionized IGM as observed today \citep{Furlanetto2016}. The evolving bubble size distribution as a function of redshift reflects how reionization progressed \citep{McQuinn2007a, Neyer2024}. Higher-order spatial statistics reveal further details about the topology, e.g., imprinting whether reionization was driven by clustered sources in overdense regions or numerous faint galaxies. However, until tomographic maps are available it is crucial to constrain the properties of ionized bubbles with indirect measurements.

One of the most promising observational probes of ionized bubbles at high redshift is the Lyman-alpha (Ly$\alpha$) emission line, which originates from the transition of electrons in hydrogen atoms from the first excited state to the ground state. Ly$\alpha$ photons are produced efficiently within star-forming galaxies due to the recombination of ionized hydrogen. 
These photons undergo a complex resonant scattering process that enhances dust absorption and alters their emergent spectral profiles \citep{Mesinger2008a, Dijkstra2019}, and at $z \gtrsim 5$ must traverse the surrounding IGM, which leads to further scattering out of the line of sight if they encounter partially neutral gas. A given galaxy embedded in a sufficiently large ionized bubble can transmit most of its Ly$\alpha$ emission, whereas in a small bubble the line is strongly attenuated \citep{Miralda-Escude1998, Santos2004, McQuinn2007b, Mesinger2008b}. As a result, the detectability of high-$z$ LAEs is sensitive to the dust content affecting escape fractions, shape of pre-IGM spectral line profiles to determine the frequency-dependent transmission, and the ionization state of the surrounding IGM including the bubble size distribution and peculiar velocities setting resonant cutoff features \citep{Furlanetto2006c, Iliev2008, Laursen2011, Dijkstra2014, Gronke2021, Park2021, Park2022, Smith2022, ChenZuyi2025}.

Ly$\alpha$ transmission is the basis for many recent studies that use LAEs as signposts of reionization. For example, the 
drop in the fraction of Lyman-break galaxies with detectable Ly$\alpha$ between $z \sim 6$ and $z \sim 7$ has been interpreted as evidence for increasing IGM neutrality and smaller host bubbles \citep{Mesinger2015, Sadoun2017, Mason2018}. More direct inferences have come from analyses of Ly$\alpha$ damping-wing absorption in spectra of galaxies and quasars \citep{Mason2020}. If a red damping wing feature is present, it signals a significant neutral column along the line of sight, whereas a sharp, unattenuated Ly$\alpha$ profile suggests a large ionized region around the source. Especially in light of recent observations from the \textit{James Webb Space Telescope} (\textit{JWST}), this dichotomy of features can be interpreted within idealized models or simulations \citep[e.g.][]{Huberty2025, Park2025, Nikolic2025} to place constraints on the reionization history of the Universe \citep{Asada2025, Mason2025, Umeda2025}. Furthermore, specific cases of very high-redshift observed galaxies have shown unexpectedly strong Ly$\alpha$ emission indicative of very large ionized bubbles \citep{Hayes2023, Endsley2022, Endsley2024, Witstok2025}, and already test models even with a few objects.

While it is clear that evolving bubble sizes influence the potential for LAE visibility, interpreting Ly$\alpha$ observations in terms of bubble sizes is non-trivial and highlights the need for further theoretical calibration of Ly$\alpha$-based probes. The simple picture of a galaxy surrounded by a spherical bubble of ionized gas can be complicated by several factors, including the intrinsic luminosity and galaxy scale radiative transfer, scattering in the circumgalactic medium and infall regions, the presence of residual neutral hydrogen within the bubble, and the 3D inhomogeneous density and velocity structure of the IGM. Many studies still employ the idealized spherical models to estimate the minimum bubble size required for a given Ly$\alpha$ transmission fraction or to infer the IGM neutral fraction from damping wing absorption profiles. While these analytic approaches yield order-of-magnitude estimates, they can suffer from degeneracies and oversimplifications \citep{Huberty2025}. In reality, the IGM during reionization is clumpy and the ionized regions have irregular, evolving shapes that can be used with a more robust synthetic observation approach to correct for model shortcomings \citep[e.g.][]{Keating2023a, Lu2025}. Comparisons of increasingly self-consistent radiative transfer modeling with real data enables us to learn potential biases that are important to account for when using LAEs to trace ionized bubbles in the EoR and the large-scale cosmology.

In this paper, we pursue such an approach using the \thesan suite of state-of-the-art radiation-hydrodynamical (RHD) simulations that of the EoR \citep{Kannan2022a, Garaldi2022, Smith2022, Garaldi2023}. \thesan combines the galaxy formation physics of the IllustrisTNG model with on-the-fly ionizing radiation transport using the \areport code \citep{Kannan2019} to simulate reionization. The flagship run has high enough resolution to accurately capture the intrinsic properties of LAEs and large enough volume to produce representative bubble sizes and clustering, so that LAE visibility and calibrated spectral features can be directly related to the known ionization structure in the simulation. Earlier results from \thesan have explored the emission and transmission statistics during reionization \citep{Smith2022}, followed by an initial calibration of the LAE luminosity function \citep[LF;][]{Xu2023}, and this paper updates that prescription to enable more robust equivalent width selection for LAE catalogs with a variety of intended applications. By connecting LAEs to the bubble size distribution this paper also serves as a follow-up study to \citet{Neyer2024}, in which we explored dependence on galaxy properties and the local environment, finding the brightest galaxies tend to reside in the largest ionized regions. Similarly, to our study of bubble merger histories in \citet{Jamieson2025}, in which we identified three stages, with early-forming bubbles growing slowly in isolation until an accelerated percolation phase (``flash ionization'') ending with rapid expansion dominated by the largest bubble, emerging by $z \approx 9$--$10$, well before the midpoint of reionization. Finally, this study is paired with an observational comparison to narrow-band selected LAE high-redshift surveys (L.A. Garc{\'\i}a et al., in prep.) and a study on detecting 21-cm--LAE cross-correlation signatures \citep{Chen2025}.

Here, our focus is on understanding what ingredients are necessary to observe LAEs, systematic connections between Ly$\alpha$ observables and bubble sizes, and tensions between simulated bubble sizes and those inferred from applying simplified analytic models to observations. A key strength of \thesan is the ability to predict intrinsic Ly$\alpha$ luminosities and IGM transmission, inherited from the realistic galaxy formation and RHD modeling. However, to accurately predict observed Ly$\alpha$ luminosities, it is necessary to account for additional small-scale processes within galaxies that are not fully resolved in large-scale simulations \citep{Khoraminezhad2025}. Therefore, an empirically calibrated model is required to incorporate radiative transfer effects due to dust absorption and resonant scattering, including the impact of outflows on the Ly$\alpha$ line profile. Still, the result is a physically-motivated framework inherited from the realistic galaxy formation and IGM modeling with predictive clustering, line-of-sight variability, and duty-cycle effects for insights into how LAEs can be used to infer the ionization state of the IGM and the progress of reionization.

The paper is organized as follows. In Section~\ref{sec:methods}, we describe the simulations and our methodology for generating synthetic Ly$\alpha$ observations and measuring ionized bubble sizes. Section \ref{sec:bestfit} presents our results from the best-fit model for observed Ly$\alpha$ properties, while Section~\ref{sec:bubbles} explores the implications for connections to ionized bubble sizes. Finally, we conclude with a summary of our key findings in Section~\ref{sec:conclusions}, also outlining directions for future work. In Appendix~\ref{appx:thesan2}, we explore the alternative physics variations within the broader \thesan simulation suite. We make the catalogs public
so others can extend the applications beyond our initial explorations.

\section{Methods}
\label{sec:methods}
In Section~\ref{subsec:thesan}, we briefly describe the \thesan simulations in general, while in Section~\ref{subsec:lya-em}, we focus on the Ly$\alpha$ emission and transmission modeling. In Section~\ref{subsec:cal}, we introduce the LAE calibration procedure.

\subsection{\thesan simulations}
\label{subsec:thesan}
The \thesan simulations are a suite of large-volume, high-resolution cosmological RHD simulations that aim to simultaneously resolve large-scale structure formation during the Epoch of Reionization (EoR) and model realistic galaxy formation processes \citep{Kannan2022a, Garaldi2022, Smith2022, Garaldi2023}. They employ the moving-mesh hydrodynamics code \areport \citep{Kannan2019}, an extension of the \arepo code \citep{Springel2010, Weinberger2020}, which incorporates radiative transfer (RT) capabilities. \areport solves the fluid equations on an unstructured Voronoi mesh that adapts to the fluid flow, enabling an accurate quasi-Lagrangian treatment of cosmological gas dynamics. The first two moments of the RT equation are solved assuming the M1 closure relation \citep{Levermore1984}, and the scheme reaches second-order accuracy. The simulations use the state-of-the-art and well-tested IllustrisTNG galaxy formation model \citep{Weinberger2017, Pillepich2018a, Springel2018}, an updated version of the original Illustris simulation framework \citep{Vogelsberger2013, Vogelsberger2014b, Vogelsberger2014a}, including detailed treatments of star formation, feedback processes, and black hole physics. Dust physics is also incorporated following \citet{McKinnon2017}. Gravity is calculated using a hybrid Tree--PM approach: long-range forces are computed using a particle-mesh algorithm, while short-range forces are calculated using a hierarchical oct-tree method. To minimize force errors during time integration, node centres are randomly shifted at each domain decomposition, as detailed in \citet{Springel2021}. The simulations also employ a hierarchical time integration scheme to efficiently handle the wide range of dynamical timescales involved.

The RT equations are coupled to a non-equilibrium thermochemistry solver to accurately compute the ionization states of hydrogen and helium, as well as temperature changes due to photo-ionization, metal cooling, and Compton cooling. A reduced speed of light approximation (RSLA) is used for computational efficiency, with an effective speed of $\tilde{c} = 0.2\,c$. Photons are discretized into energy bins defined at thresholds: $[13.6, 24.6, 54.4, \infty)\,\text{eV}$. Each cell in the simulation tracks the comoving photon number density and flux for each bin, with the luminosity sourced by star particles based on the age and metallicity (although mixed transport coefficients use a 2\,Myr, $0.25\,\Zsun$ stellar population), computed using the Binary Population and Spectral Synthesis models \citep[BPASS;][]{Eldridge2017,Stanway2018}, assuming a Chabrier IMF \citep{Chabrier2003}. An additional birth cloud escape fraction parameter, $f_\text{esc}^\text{cloud} = 0.37$, is added to mimic the absorption of LyC photons happening below the resolution scale of the simulation, tuned to reproduces a realistic late-reionization history.

The \thesan simulations have a box size of $L_\text{box} = 95.5\,\text{cMpc}$, assuming cosmological parameters from \citet{Planck2016} with $h = 0.6774$, $\Omega_m = 0.3089$, $\Omega_\Lambda = 0.6911$, and $\Omega_b = 0.0486$. The flagship high-resolution simulation, \thesanone, contains $2100^3$ dark matter and (initial) gas particles for resolutions of $m_\text{DM} = 3.12 \times 10^6\,\Msun$ and $m_\text{gas} = 5.82 \times 10^5\,\Msun$, respectively, effectively resolving atomic cooling haloes by at least $50$ particles run down to $z = 5.5$. This allows for a self-consistent exploration of how ionized bubbles evolve around galaxies in response to their emission of ionizing radiation. We also analyze alternative physics runs with eight times lower mass resolution (discussed when relevant). The scientific scope of the \thesan simulations spans from the physics of ionizing sources to large-scale IGM observables. Prior work has already established the role of ionizing escape fractions \citep{Yeh2023} and local environments \citep{Zhao2025} on reionization and explored in detail the evolution of individual high-$z$ galaxies \citep[][in \textsc{thesan-zoom}]{Kannan2025}. Within this context, our results offer a new perspective on galactic Ly$\alpha$ emission as a probe of ionized bubbles, synergizing with existing studies of how high-$z$ galaxy properties impact and track the state of the surrounding IGM \citep{Garaldi2024a, Garaldi2024b, Kakiichi2025} and complementary bubble size probes like emission line intensity mapping \citep{Kannan2022b, Almualla2025}.

\subsection{Ly$\alpha$ emission and transmission}
\label{subsec:lya-em}
A key strength of the \thesan simulations is their ability to predict intrinsic Ly$\alpha$ emission and transmission properties from high-redshift galaxies, which is crucial for understanding observational probes of reionization. The Ly$\alpha$ emission line, produced by $2p \rightarrow 1s$ transitions of hydrogen atoms, is sensitive to the local ionization state of the surrounding gas. Galaxies in \thesan exhibit a wide range of intrinsic Ly$\alpha$ luminosities, modulated by factors such as the star-formation rate. The details for calculating the total intrinsic emission due to recombinations, collisional excitation, and local stars are described in \citet[][see their equations 1--3]{Smith2022}, which we summarize in a single equation as:
\begin{equation} \label{eq:L_stars}
  \frac{L_{\alpha}}{h \nu_\alpha} = \int \left[ P_\text{B} \alpha_\text{B} n_p + q_\text{col} n_\text{\HI} \right] n_e \text{d}V + 0.68 (1 - f_\text{esc}^\text{cloud}) \dot{N}_\text{ion} \, .
\end{equation}
Here $h \nu_\alpha = 10.2\,\text{eV}$, $P_\text{B}$ is the Ly$\alpha$ conversion probability per recombination event, $\alpha_\text{B}$ is the case B recombination coefficient, $q_\text{col}$ is the collisional rate coefficient, and $n_p$, $n_\text{\HI}$, and $n_e$ are number densities for protons, neutral hydrogen, and electrons. Finally, $0.68$ is the fiducial conversion probability, $f_\text{esc}^\text{cloud}$ denotes the escape fraction of ionizing photons calibrated for each simulation to match reionization history constraints, and $\dot{N}_\text{ion}$ is the age and metallicity dependent emission rate of ionizing photons from stars taken from the BPASS models (v2.2.1). Combining resolved and sub-resolution emission mechanisms is necessary given that $f_\text{esc}^\text{cloud} < 1$. Moreover, this approach allows us to utilize the on-the-fly ionization field while also reproducing the expected ionizing photon budget.

Furthermore, the simulations provide robust predictions of the frequency-dependent transmission of Ly$\alpha$ photons through the IGM from snapshots corresponding to $z = \{5.5, 6, 6.6, 7, 8, 9, 10, 11, 13\}$, which is essential for linking intrinsic Ly$\alpha$ luminosities to the observed signals by accounting for scattering out of the sightline due to intervening neutral hydrogen. The data is available for a large number $\approx 90$ per cent of the most massive centrals, extracted for 768 radially outward rays corresponding to the equal area healpix directions of the unit sphere with the \colt code \citep{Smith2015,Smith2019,Smith2022MW} for exact ray tracing through the native Voronoi unstructured mesh data. The rays start at initial distances of $R_\text{vir}$, defined as the radius within which the mean density becomes 200 times the cosmic value ($R_{200}$), taking the systemic location and rest-frame from the subhalo intrinsic Ly$\alpha$ luminosity averaged position $\bm{r}_\alpha$ and velocity $\bm{v}_\alpha$. The transmission is calculated over a broad wavelength range of $\Delta v \in [-2000, 2000]\,\text{km\,s}^{-1}$ sampled at a high spectral resolution of $5\,\text{km\,s}^{-1}$ or a resolving power of $R \approx 60\,000$, integrating out to a distance of $4000\,\text{km\,s}^{-1} / H(z) \approx 40\,\text{cMpc}\,[(1+z)/7]^{-1/2}$. Optical depths are based on the continuous Doppler shifting scheme described in \citet{Smith2022}, which incorporates velocity gradients encountered during propagation. The total optical depth $\tau$ defines the frequency-dependent transmission function:
\begin{equation}
  \mathcal{T}_\text{IGM}(\Delta v) \equiv \exp\big[ -\tau(\Delta v) \big] \, ,
\end{equation}
describing the fraction of flux not attenuated by the IGM after escaping the halo. Additional damping-wing absorption $\tau_\text{DW}$ from the distant IGM beyond the local rays is included in a statistical sense based on the global reionization history, including the complete first-order quantum-mechanical correction to the Voigt profile presented by \citet{Lee2013}, strengthening the red wing due to positive interference of scattering from all other levels.

While \thesan can accurately model intrinsic Ly$\alpha$ luminosities and IGM transmission, detailed LAE modeling is strongly affected by interstellar medium (ISM) scale sourcing and radiation transport through the cirgumgalactic medium (CGM). The uncertainties mainly arise from the galaxy formation model (and limited resolution), which includes temporary decoupling of wind particles from the hydrodynamics and the use of an effective equation of state (EoS) for cold gas above the density threshold $n_\text{H} \approx 0.13\,\text{cm}^{-3}$ \citep{SpringelHernquist2003}. The observed Ly$\alpha$ from galaxies also depend on small-scale processes that are not fully resolved in the large-scale \thesan simulations and require empirical models to account for their effects. Dust, in particular, can significantly attenuate Ly$\alpha$ emission, while galactic outflows and winds can alter the escape paths and emergent spectra of Ly$\alpha$ photons, directly affecting the subsequent IGM reprocessing. In fact, the bright end of the observed Ly$\alpha$ LF differs from the intrinsic one by up to two orders of magnitude. Thus, we choose to apply a calibrated model to incorporate these processes and inform us about what is required to match observations.

In this paper, we present improved LAE catalogs from \thesan, building on the efforts of \citet{Xu2023}, which we leverage to investigate the relationship between LAE properties and the sizes of the surrounding ionized regions. We also explore how Ly$\alpha$ luminosities, escape fractions, equivalent widths, and transmission vary with the host galaxy's mass, star-formation rate, and local environment, including how these trends evolve as reionization progresses. This offers insights into how LAEs can be used to infer ionized bubble sizes and the reionization state of the IGM during the EoR.

\subsection{Ly\,$\alpha$ calibration procedure}
\label{subsec:cal}
We now describe the empirical calibration enabling a direct comparison of synthetic Ly$\alpha$ properties predicted by \thesan to those observed in our Universe. As discussed in \citet{Xu2023}, most previous models applied to the EoR do not account for dust in addition to IGM transmission, which strongly suppresses more massive galaxies that are intrinsically bright but have low Ly$\alpha$ escape fractions. We must simultaneously constrain the impacts of dust, winds, and other processes that are not fully resolved in the simulations. Compared to \citet{Xu2023}, we incorporate a more direct wind model and the dust absorption is self-consistent with the stellar continuum. Finally, we report the parameters to best fit the observed Ly$\alpha$ LF.

We choose our emergent spectral model to be the approximate analytic solution for a Ly$\alpha$ point source within a spherical cloud undergoing slow expansion or contraction, which produces an asymmetric double-peaked profile \citep{Nebrin2025, Smith2025}. This is physically motivated, naturally scales the line width with the peak separation, and allows both blue and red dominated spectra. We relate the velocity offset to the dimensionless frequency:
\begin{equation} \label{eq:x}
  x \equiv \frac{\nu - \nu_0}{\Delta \nu_\text{D}} = -\frac{\Delta v}{v_\text{th}} \, ,
\end{equation}
where $\Delta v$ denotes the velocity offset, $\nu_0 = 2.466 \times 10^{15}\,\text{Hz}$ the frequency at line centre, $\Delta \nu_\text{D} \equiv (v_\text{th}/c)\nu_0$ the Doppler width of the profile, and $v_\text{th} \equiv (2 k_\text{B} T / m_\text{H})^{1/2} \approx 12.85\,T_4^{1/2}\,\text{km\,s}^{-1}$ the thermal velocity, defining the normalized temperature $T_4 \equiv T / (10^4\,\text{K})$. The `damping parameter', $a \equiv \Delta \nu_L /2 \Delta \nu_D \approx 4.702 \times 10^{-4} T_4^{-1/2}$, describes the relative broadening compared to the natural line width $\Delta \nu_\text{L} = 9.936 \times 10^7\,\text{Hz}$. In isothermal conditions the optical depth at line centre is $\tau_0 = N_\text{\HI} \sigma_0$ with neutral hydrogen column density $N_\text{\HI}$ and cross-section $\sigma_0 = f_{12} \sqrt{\pi} e^2 / (m_e c \Delta \nu_\text{D}) \approx 5.898 \times 10^{-14}\,T_4^{-1/2}\,\text{cm}^2$ and oscillator strength of $f_{12} = 0.4162$. The spectral profile is proportional to
\begin{equation} \label{eq:J_x}
  J(x) \propto x^2\,\text{sech}^2 \left(\sqrt{\frac{\pi^3}{54}} \frac{x^3}{a\tau_0}\right) \, \exp\left(-\frac{\sqrt{\pi}}{3}\frac{\beta x^3}{a\tau_0}\right) \, ,
\end{equation}
such that the model is fully described by a reference column density $N_\text{\HI}$, temperature $T$, and strength of the wind\footnote{Equation~(\ref{eq:J_x}) is only accurate for slow expansion ($\beta > 0$) or contraction ($\beta < 0$) and can only be normalised if $|\beta| < \pi \sqrt{6} / 3$. However, we use this as an effective model over a finite frequency range so we do not restrict $|\beta|$.} $\beta \equiv V_\text{max} / v_\text{th}$ where $V_\text{max}$ is the velocity at the edge of the imaginary homologous system. We are able to parameterise the first two with respect to a reference peak velocity offset $\Delta v_\text{peak}$. We adopt the static value for the peak locations, $x_\text{p} = \pm f_\text{p} (a \tau_0)^{1/3}$, where in this case $f_\text{p} = 0.93099$ \citep{Lao2020}. Under an arbitrary normalization chosen to be consistent with observations, $\Delta v_\text{peak,300} \equiv \Delta v_\text{peak} / (300\,\text{km\,s}^{-1})$, the effective optical depth is $a \tau_0 = (\Delta v_\text{peak} / f_\text{p} v_\text{th})^3 \approx 15787\,T_4^{-3/2}\,\Delta v_\text{peak,300}^3$ and the reference column density is $N_\text{\HI} \approx 5.6928 \times 10^{20}\,\text{cm}^{-1}\,T_4^{-1/2}\,\Delta v_\text{peak,300}^3$. In the end we arrive at a convenient expression for the spectral profile in terms of a peak normalized velocity offset $\Delta \tilde{v} \equiv \Delta v / \Delta v_\text{peak}$,
\begin{equation} \label{eq:J_v}
  J(\Delta \tilde{v}) \propto \Delta \tilde{v}^2\,\text{sech}^2 \left(\sqrt{\frac{\pi^3}{54}} f_\text{p}^3 \Delta \tilde{v}^3\right) \, \exp\left(\frac{\sqrt{\pi}}{3} \beta f_\text{p}^3 \Delta \tilde{v}^3\right) \, .
\end{equation}
We parameterise the peak position and wind strength according to
\begin{equation} \label{eq:Dv_peak}
  \Delta v_\text{peak,300} = a_\text{peak}\,V_{c,300} + b_\text{peak}
\end{equation}
and
\begin{equation} \label{eq:beta_wind}
  \beta = a_\text{wind}\,V_{c,300} + b_\text{wind} \, ,
\end{equation}
where $V_c$ is the circular velocity of the galaxy and $V_{c,300} \equiv V_c / (300\,\text{km\,s}^{-1})$, while the $a_\text{peak}$/$a_\text{wind}$ coefficients and $b_\text{peak}$/$b_\text{wind}$ constant offsets are free parameters. This updated model is in line with the expectation that more massive galaxies have larger velocity offsets \citep{Yang2016,Verhamme2018}. We ignore the normalization constant since it will cancel out later when calculating the fraction of transmitted luminosity.

To calculate this transmission ratio, the line profile from each galaxy is multiplied by the frequency-dependent IGM transmission fraction of the central galaxy in its group. We expect satellites to have similar IGM transmission with the exception of minor localized sightline and velocity offset effects. We integrate over frequency and then divide by the integral of the intrinsic profile:
\begin{equation}
  \mathcal{T}_\text{IGM} = \frac{\int \text{d}\nu\,J(\nu) e^{-\tau(\nu)}}{\int \text{d}\nu\,J(\nu)} \, .
\end{equation}
We then multiply $\mathcal{T}_\text{IGM}$ by the escape fraction from dust, $f_\text{esc}$, which depends on the parameters of our dust model discussed next. This gives the final observed transmission fraction:
\begin{equation} \label{eq:fescTigm}
  f_\text{esc} \times \mathcal{T}_\text{IGM} = L_{\rm Ly\alpha}^\text{obs} / L_{\rm Ly\alpha}^\text{int} \, .
\end{equation}
From this, we calculate the observed Ly$\alpha$ luminosity, which gives us the observed LFs to calibrate to observations.

We update the dust model to ensure that the Ly$\alpha$ escape fraction is less than the escape fraction of the non-resonant stellar continuum around the line (1216\,\AA). We emphasize that dust corrections are particularly important for the high-luminosity end of the UV and Ly$\alpha$ LFs. However, the amount of dust and its composition is still somewhat uncertain at these high redshifts. Therefore, we use an empirical dust-attenuation ($A_{1500}$) model, which is obtained by fitting the IRX--UV relationship inferred from ALMA observations at $z\sim4$--$7$ in \citet{Bouwens2016} to the following equation:
\begin{equation} \label{eq:AUV}
  A_{1500} = 2.5\,\log\left(1 + 10^{0.4\alpha_\text{dust}(M_\text{dust}(z) - M_\text{1500,int})}\right) \, ,
\end{equation}
where
  $M_\text{dust}(z) = M_\text{dust,0} + z \cdot M_{\text{dust},z} \, ,$
and we follow the same parametrization as \citet{Kannan2023} with $\alpha_\text{dust} = 0.5133$, $M_{\text{dust},z} = -0.12504$, and $M_\text{dust,0} = -20.61$ for UV measurements taken at 1500\,\AA. 
This definition for attenuation satisfies a convenient relation translating between intrinsic and observed UV magnitudes: 
  $M_{1500} = M_\text{1500,int} + A_{1500}$ \citep{Behroozi2020}.
Furthermore, we can connect back to the attenuation optical depth ($\tau_{1500}$, as parametrized in \citealt{Vogelsberger2020b}) and hence the escape fraction
via
  $f_\text{esc}^{1500} = L_{1500}/L_\text{1500,int} = \exp\left(-\tau_{1500}\right) \, ,$
such that $A_{1500} = 2.5\,\tau_{1500} / \ln(10) \approx 1.086\,\tau_{1500}$. 
For our model, we require self-consistent Ly$\alpha$ and continuum dust corrections and further require that the Ly$\alpha$ escape fractions are no larger than the continuum ones. Specifically, Eq.~(\ref{eq:AUV}) provides the attenuation at {1500\,\AA}, which we can relate to the value at {1216\,\AA} via the wavelength dependence of the UV dust opacity $\propto \lambda^{-1.1}$ 
, such that
  $A_{1216}/A_{1500} = \tau_{1216}/\tau_{1500} \approx \left(1500\,\text{\AA}/1215.67\,\text{\AA}\right)^{1.1} \approx 1.26 \, $
\citep{Gnedin2014}.
Therefore, in our updated model we limit the Ly$\alpha$ escape fraction with the following parameterization
\begin{equation} \label{eq:fescA}
  f_\text{esc}^\alpha = f_\text{esc}^{1216}\,\exp\left( -a_\text{dust} A_{1216}^{b_\text{dust}} \right) \, ,
\end{equation}
where $a_\text{dust}$ and $b_\text{dust}$ control the cutoff scale and speed. We again emphasise that this is chosen to allow both escape fractions to go from unity for faint galaxies (negligible absorption) and asymptotically approach zero for high mass galaxies where we expect much lower equivalent widths \citep[see the discussion in][]{Xu2023}.

Unfortunately, competing physical effects in these models can become degenerate; for example, strong dust absorption lowers the observed Ly$\alpha$ luminosity, but so will a spectral model with insufficient flux redward of line centre. Despite these uncertainties, it is a well-defined optimisation problem to calibrate our model to observed LFs. For a given model, we construct the LF combining all 768 lines of sight to increase the statistical power at the bright end. With efficient code we are able to evaluate $10^5$ models arranged as a Latin hypercube within the six-dimensional parameter space.

We compare our simulated LFs with the observations from \citet{Ouchi2008, Ouchi2010, Konno2018}. We do not use the upper limit from \citet{Ouchi2010} as it is consistent with the data points from \citet{Konno2018}. For each of the observed LFs, we evaluate our models with the corresponding evenly spaced bins and bin centers, so that the penalty functions are as consistent and reliable as possible. We calculate the raw penalties ($P_z$), using the LF calculated using the model ($\phi_{\rm m}$), the data points of the observed LF ($\phi_{\rm o}$), and the one-sigma errors, $\sigma_{\rm m}$ and $\sigma_{\rm o}$ in $\log \phi_{\rm m}$ and $\log \phi_{\rm o}$, respectively:
\begin{equation} \label{eq:pen}
  P_z = 1 - \exp\left[-\frac{1}{2} \left(\log \phi_{\rm m} - \log \phi_{\rm o}\right)^2 / \left(\sigma_{\rm m}^2 + \sigma_{\rm o}^2\right)\right] \, .
\end{equation}
We find the best-fit parameters by minimizing the total penalty:
\begin{equation}
\label{eq:tot_pen}
    P_{\rm tot} = \sum_{z} \frac{P_z}{n_{\text{overlap,}z} \times n_{\text{obs,}z}} \, ,
\end{equation}
which for each set of parameters is the sum of the raw penalties for each redshift ($P_z$) divided by the number of overlapping bins ($n_{\text{overlap,}z}$) and the number of observational points for that redshift ($n_{\text{obs,}z}$). The number of overlapping bins is defined as the number of observed bins with more than one galaxy according to the model. This ensures that our best-fit parameters will lead to nonzero LFs for as many of the luminosity bins as possible, while other criteria sometimes match well for certain bin ranges at the expense of the overall fit. Thus, we expect to find fits extending across the entire range from faint to bright LAEs, as those populations should be covered by the large volume and high resolution of \thesan. The final values of the best-fit parameters determined by minimizing this total penalty are shown in Table~\ref{tab:bestfit}.

\begin{table}
    \centering
    \caption{Best-fit parameters, found by simultaneously fitting our models for $z=5.5$ and $z=6.6$ to observational data of Ly$\alpha$ luminosity functions from \citet{Ouchi2008, Ouchi2010, Konno2018}.}
    \addtolength{\tabcolsep}{3pt}
    \renewcommand{\arraystretch}{1.3}
    \begin{tabular}{cccccc}
        \hline
        $a_{\rm peak}$ & $\log b_{\rm peak}$ & $a_{\rm wind}$ & $b_{\rm wind}$ & $a_{\rm dust}$ & $b_{\rm dust}$\\
        \hline
        0.011 & $-0.440$ & $-0.088$ & 1.399 & 0.007 & 1.902\\
        \hline
    \end{tabular}
    \addtolength{\tabcolsep}{-3pt}
    \label{tab:bestfit}
    \vspace{0.5\baselineskip}
\end{table}

\begin{figure*}
    \centering
    \includegraphics[scale=1]{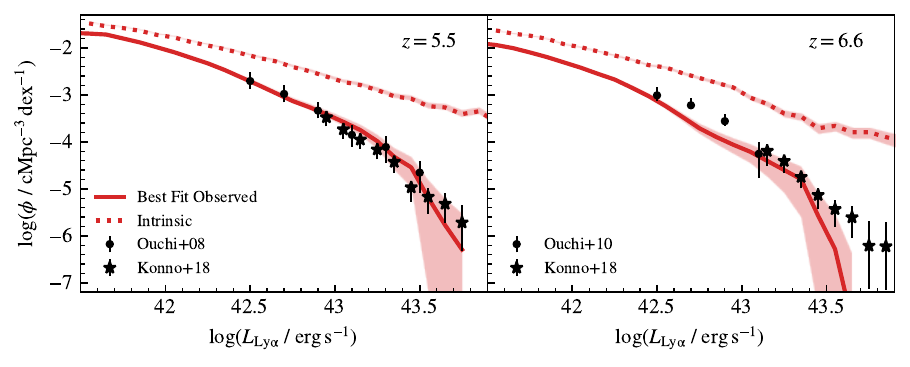}
    \caption{Ly$\alpha$ luminosity functions (red) using the best-fit model calibrated jointly for $z=5.5$ and $z=6.6$, shown with the black points. Note that the data points in the left panel are observed at $z=5.7$ and we compare to $z=5.5$ in \thesan, as we expect this difference to be small. The 16$^\text{th}$ to 84$^\text{th}$ percentile ranges are shown as the red shaded regions. 
    The LFs and uncertainties from the ``intrinsic''  Ly$\alpha$ luminosities, which include stellar emission and IGM transmission, but not ISM- or CGM-scale dust-attenuation, are shown in the dotted curves and surrounding shaded regions. We find good agreement between our best-fit model and the observed data points for both redshifts, with some underestimation of the LFs for the $z=6.6$. The discrepancy between $10^{42.5}$ and $10^{43}\,\rm erg\,s^{-1}$ is likely due to the limitations of the model and simultaneously fitting both redshifts. The underestimation at the high-luminosity end is due to box size effects preventing creation of the very rare bright sources.}
    \label{fig:bestfitLF}
\end{figure*}

\begin{figure*}
    \centering
    \includegraphics[width=\linewidth]{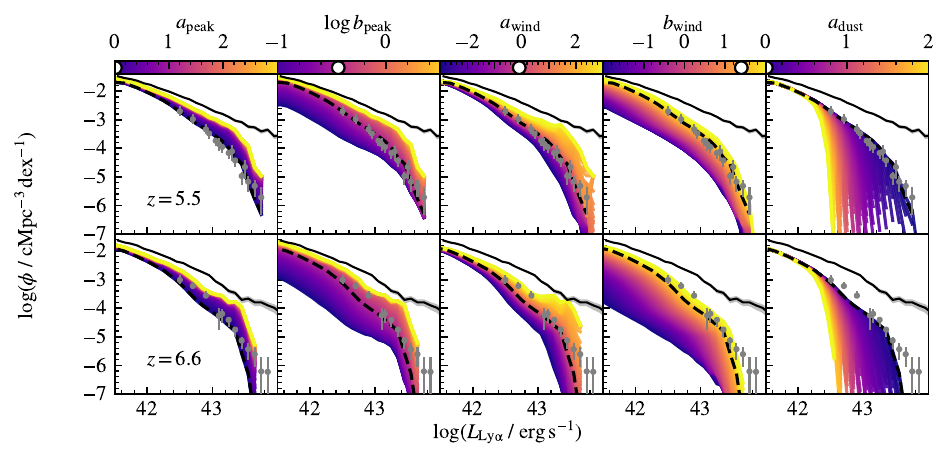}
    \caption{Parameter sweeps for $z=5.5$ (upper panels) and $z=6.6$ (lower panels) around the best fit parameters listed in Table~\ref{tab:bestfit}. Each panel shows the observed luminosity functions calculated by sweeping each parameter over the allowed range while holding all other parameters fixed at their best-fit values. Each parameter uniquely affects the shape and normalization of the resulting LFs, with $b_{\rm dust}$ having little effect around the best-fit values for the other parameters.}
    \label{fig:sweeps}
\end{figure*}

\section{LAEs from the best-fit model}
\label{sec:bestfit}
Here, we show results from the best-fit model. We apply the best-fit parameter values from Table~\ref{tab:bestfit} to calculate the Ly$\alpha$ escape fraction ($f_\text{esc}^\alpha$) and IGM transmission ($\mathcal{T}_\text{IGM}$) for each of the galaxies with intrinsic Ly$\alpha$ luminosity above $10^{41.5}\, \rm erg\,s^{-1}$. In this work, we use ``intrinsic'' to refer to the uncalibrated Ly$\alpha$ luminosity calculated directly from the stellar-scale emission and IGM transmission, but without accounting for attenuation due to dust in the ISM and CGM. The IGM transmission is calculated along each of the 768 lines of sight from each of the galaxies, whereas the escape fractions are calculated for each galaxy. We create catalogs of $f_\text{esc}^\alpha$, $\mathcal{T}_\text{IGM}$, and the resulting observed Ly$\alpha$ luminosities along each line of sight for each of the galaxies calculated using Eq.~(\ref{eq:fescTigm}). In these catalogs, we also consolidate observed UV magnitudes ($M_{1500}$), Ly$\alpha$ equivalent widths (EW=$L_{\rm Ly \alpha}^{\rm obs}/(f_\text{esc}^{1216}\,L_{\lambda,1216})$), and other quantities such as MFP bubble sizes along each line of sight ($R_\text{LOS}$) and environmental overdensity ($\delta = \rho / \bar{\rho} - 1$) for convenience.

\subsection{Luminosity functions}
\label{subsec:thesan1}

The synthetic observed Ly$\alpha$ luminosities calculated using the best-fit parameters are used to make luminosity functions at redshifts $z = 6.6$ and $z = 5.5$ as shown in Figure~\ref{fig:bestfitLF}. The data points from from \citet{Ouchi2008, Ouchi2010, Konno2018}, shown in black, indicate measurements of the Ly$\alpha$ LFs at these redshifts. The $z=5.5$ comparison is made with observations at $z=5.7$, but the small redshift difference is expected to have a small effect on the results. The red line traces the modeled Ly$\alpha$ LF, representing the median prediction of the model at both redshifts, with the red shaded region marking the 16$^\text{th}$ to 84$^\text{th}$ percentile ranges, providing an indication of the statistical uncertainty in the model. In cases where the Poisson noise for the model is as large as the median value of the LF, 2 dex error bars are applied for fitting to avoid numerical errors, and to de-prioritize bins with very few modeled galaxies. Additionally, the red dotted line represents the intrinsic Ly$\alpha$ LF, which is calculated in the simulation self-consistently, and similarly shows the 16$^\text{th}$ to 84$^\text{th}$ percentile ranges with red shaded regions. The difference between the dotted (intrinsic) and solid (observed) LFs illustrates the impact of the IGM and small-scale processes, such as dust and outflows, in reducing the observable Ly$\alpha$ luminosity. 

The observed LFs shown in Figure~\ref{fig:bestfitLF} demonstrate both the accuracy of the model in reproducing the observed LFs and the importance of accounting for attenuation mechanisms when interpreting Ly$\alpha$ observations. These updated LFs are slightly lower than those presented in \citet{Xu2023}, reflecting differences in the model parameterization and calibration data sets. The discrepancy between the updated LFs and observed LF data points for $z=6.6$ at the fainter end are likely due to the effects of simultaneously fitting to the $z=5.5$ and $z=6.6$ data. The fit is dominated by the $z=5.5$ points due to the higher number of data points and the lower uncertainties. For the LF at $z=6.6$, our model underpredicts bright sources due to box size effects. The \thesan box is not large enough to robustly produce the rare, very bright sources probed by the high-luminosity end of the \citet{Konno2018} data.

To visualize the impact of the fitting parameters on the LF, we perform parameter sweeps, as shown in Figure~\ref{fig:sweeps} for redshifts $z = 5.5$ (upper panels) and $z = 6.6$ (lower panels). In this analysis, we systematically vary the key parameters, $a_{\rm peak}$, $b_{\rm peak}$, $a_{\rm wind}$, $b_{\rm wind}$, and $a_{\rm dust}$, to explore their individual effects on the predicted LFs. These parameters represent the peak position of the Ly$\alpha$ emission line ($a_{\rm peak}$, $b_{\rm peak}$), the influence of galactic winds on Ly$\alpha$ photon escape ($a_{\rm wind}$, $b_{\rm wind}$), and the impacts of dust attenuation ($a_{\rm dust}$, $b_{\rm dust}$). Note that changing $b_{\rm dust}$ has a negligible effect on the LF at the best-fit point in parameter space. Each plot shows how varying one of these parameters through its allowed range, while holding all other parameters fixed at their best-fit values (as provided in Table~\ref{tab:bestfit}) affects the resulting LF. This allows us to isolate the contribution of each parameter to the shape and normalization of the Ly$\alpha$ LF. 

The peak position parameters, $a_{\rm peak}$ and $b_{\rm peak}$, describe the location of the peak of the Ly$\alpha$ emission line. This serves to characterize the peak velocity offset which plays a crucial role in determining how Ly$\alpha$ photons propagate by determining how much of the Ly$\alpha$ emission will be shifted out of resonance leading to higher transmission. Varying these parameters alters both the normalization and the slope of the LF, with lower values of $a_{\rm peak}$ and $b_{\rm peak}$ indicating less of a velocity offset and decreasing the resulting observed luminosities. The wind parameters, $a_{\rm wind}$ and $b_{\rm wind}$, capture the effects of galactic outflows on the escape fraction of Ly$\alpha$ photons. Galactic winds can either enhance or inhibit the transmission of Ly$\alpha$ photons by redistributing neutral hydrogen in the galaxy's circumgalactic medium. Sweeping over these parameters highlights their impact on the LF, especially in the intermediate to high-luminosity range where the shape of the function becomes most sensitive to $a_{\rm wind}$. The dust parameters, $a_{\rm dust}$ and $b_{\rm dust}$, govern how dust extinction affects Ly$\alpha$ photons. Specifically, $a_{\rm dust}$ has a noticeable impact on the observed LF, particularly in the steepness of the drop off at the bright luminosity end. In contrast, $b_{\rm dust}$ shows little to no effect near the best-fit values of other parameters, indicating that it is relatively unconstrained and plays a minor role in determining the overall shape of the LF. This suggests that while dust attenuation is important, it is predominantly governed by $a_{\rm dust}$, with minimal influence from $b_{\rm dust}$.

\subsection{Dependence on UV brightness}

\begin{figure*}
\centering
    \includegraphics[width=\linewidth]{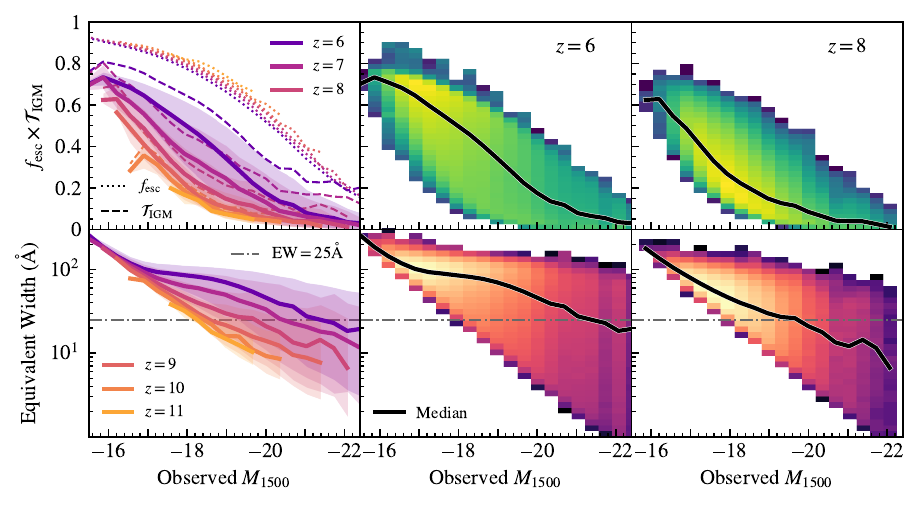}
    \caption{
    Upper panels: Product of escape fraction ($f_{\rm esc}$) and IGM transmission ($\mathcal{T}_{\rm IGM}$) versus observed UV magnitude ($M_{1500}$). Lower panels: Ly$\alpha$ equivalent width (EW) versus $M_{1500}$. Left column: Medians across redshifts (colored curves), with shaded 16th–84th percentile ranges. For \fT, dotted and dashed lines also show individual $f{\rm esc}$ and $\mathcal{T}_{\rm IGM}$ components. Center and right columns: 2D distributions at $z=6$ and $z=8$, with black lines marking medians. Brighter galaxies tend to show lower \fT and EWs, consistent with stronger dust attenuation and winds in more massive systems. All galaxies have observed Ly$\alpha$ luminosities above $10^{41.5}$ erg s$^{-1}$.
    }
    \label{fig:2dfescTigm_EW}
\end{figure*}

\begin{figure}  
\centering
    \includegraphics[width=\linewidth]{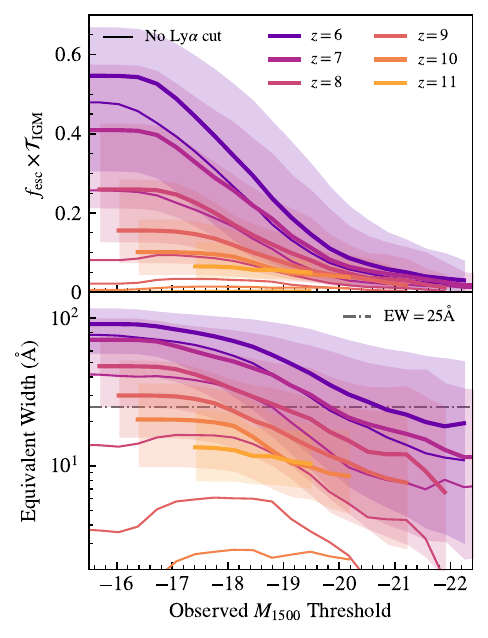}
    \caption{Upper panel: Total transmission \fT from galaxies brighter than the observed UV magnitude ($M_{1500}$) threshold. Lower panel: Equivalent width of the Ly$\alpha$ line versus $M_{1500}$ threshold. The dashed gray line indicates the 25\,\r{A} equivalent width threshold for Ly$\alpha$ emitters (LAEs). In both panels, the thick curves show the medians for galactic lines of sight with observed Ly$\alpha$ luminosity above $10^{41.5}\,\rm erg\,s^{-1}$, with shaded regions indicating the 16$^\text{th}$ to 84$^\text{th}$ percentile ranges. The thin curves show the medians for all of the galaxies and lines of sight included in this study with only a cut on the intrinsic Ly$\alpha$ luminosity cut at $10^{41.5}\,\rm erg\,s^{-1}$. Brighter galaxies generally exhibit lower Ly$\alpha$ equivalent widths and escape fractions, reflecting the more significant dust attenuation and wind effects in more massive systems.}
    \label{fig:fescTigm_EW}
\end{figure}
The observability of LAEs depends on the UV brightness, characterized by the UV magnitude at 1500\,\AA\xspace($M_{1500}$). Thus, we explore the connection between the calibrated Ly$\alpha$ properties and observed UV magnitudes using the updated dust prescription. Figure~\ref{fig:2dfescTigm_EW} shows two key relationships derived from the \thesanone simulation: the product of the Ly$\alpha$ escape fraction and IGM transmission as a function of observed UV magnitude (upper panels), and the Ly$\alpha$ rest-frame equivalent width (EW) as a function of observed UV magnitude (lower panels). The medians and 16$^\text{th}$ to 84$^\text{th}$ percentile ranges are presented for a range of redshifts in the leftmost panels, with two-dimensional histograms shown in the center and rightmost panels.

The product of the Ly$\alpha$ escape fraction, $ f_{\rm esc} $, and the intergalactic medium (IGM) transmission, $ \mathcal{T}_{\rm IGM} $ captures the combined effect of Ly$\alpha$ photon escape from galaxies and subsequent transmission through the surrounding IGM. At lower UV magnitudes (corresponding to more luminous galaxies), the product of \fT tends to be lower, reflecting the increased dust attenuation and wind effects in larger galaxies that can hinder the escape of Ly$\alpha$ photons. In contrast, at fainter magnitudes, the product increases, as smaller galaxies are less affected by dust and winds, allowing more Ly$\alpha$ photons to escape and propagate through the IGM. This trend is evident across all redshifts, with the overall range increasing with decreasing redshift, as the IGM becomes increasingly ionized, leading to lower Ly$\alpha$ absorption. This behavior highlights the evolving IGM conditions as the universe reionizes. 

Similarly, the EW is a measure of the strength of the Ly$\alpha$ emission relative to the underlying stellar continuum, with higher values indicating stronger Ly$\alpha$ emission. As with \fT, UV-bright galaxies show lower EWs due to increased attenuation from galactic dust and winds. LAEs, often defined as having Ly$\alpha$ EW > 25\,\AA, are more prevalent in fainter galaxies (higher $M_{1500}$) and at lower redshifts.

Since observations are generally biased toward brighter galaxies (either in rest-frame UV or Ly$\alpha$), we investigate the impacts of observed $M_{1500}$ and observed \LLya on the characteristic Ly$\alpha$ \fT and EW. Fig.~\ref{fig:fescTigm_EW} shows the characteristic \fT (upper panel) and EW (lower panel) in galaxies with observed UV magnitude brighter than the $M_{1500}$ threshold indicated on the horizontal axis. The thick curves and shaded regions show medians and 16$^\text{th}$ to 84$^\text{th}$ percentile ranges for galaxies with observed Ly$\alpha$ luminosity above $10^{41.5}\ \rm erg\,s^{-1}$, while the thin curves are medians for all galaxies in the catalog (with intrinsic $L_{\rm Ly\alpha} > 10^{41.5}\ \rm erg\,s^{-1}$). As expected, galaxies brighter in the UV have lower \fT and EW. Including galaxies with highly attenuated Ly$\alpha$ reduces characteristic \fT and EW overall, and especially for fainter galaxies in the UV. Early in reionization, the \fT and EW for UV-faint galaxies are generally low due to the impact of residing primarily in small bubbles, which strongly attenuate the emitted Ly$\alpha$ photons. Thus, when a cut in \LLya is applied to reflect observability constraints, many of these faint systems are left out, leading to a bias toward higher \fT and EW, as shown in the difference between the thick and thin curves in Fig.~\ref{fig:fescTigm_EW}.

Due to the difficulty of observing galaxies' Ly$\alpha$ properties, we impose a cut of observed \LLya $> 10^{41.5}\ \rm erg\,s^{-1}$ for the rest of our analysis, unless otherwise noted. This cut also avoids bias in the population statistics resulting from the initial catalog selection of intrinsic \LLya $> 10^{41.5}\ \rm erg\,s^{-1}$.

\begin{figure}
    \centering
    \includegraphics[width=\linewidth]{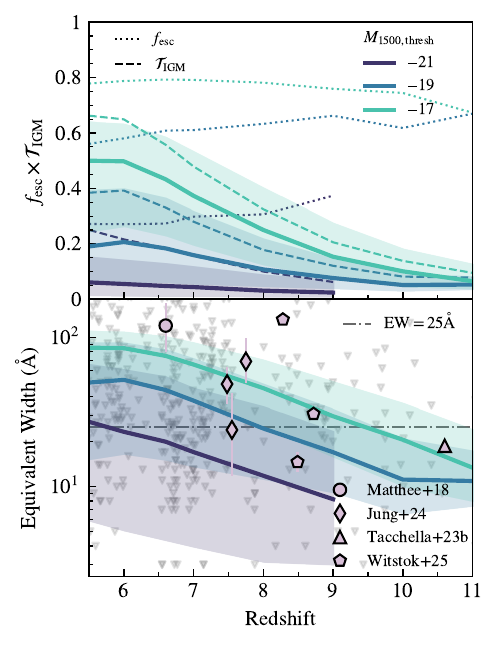}
    \caption{Upper panel: Median transmission \fT (solid) versus redshift for galaxies brighter than the UV magnitude ($M_{1500}$) thresholds indicated by color. Shaded regions show the 16$^\text{th}$ to 84$^\text{th}$ percentiles, while dotted and dashed curves represent $f_{\rm esc}$ and $\mathcal{T}_{\rm IGM}$, respectively. Lower panel: Ly$\alpha$ equivalent width (EW) versus redshift for the same magnitude bins. The grey dash-dotted line marks the common LAE threshold at 25\,\AA. The pink points show EWs from observations of individual sources, while the grey triangles show upper limits from \citet{Kageura2025, Jones2025, Tang2024b}, which are generally consistent with our models, with some potential bias toward stronger Ly$\alpha$ in observed LAEs at lower redshifts.}
    \label{fig:fescTigm_z}
\end{figure}

\begin{figure}
    \centering
    \includegraphics[width=\columnwidth]{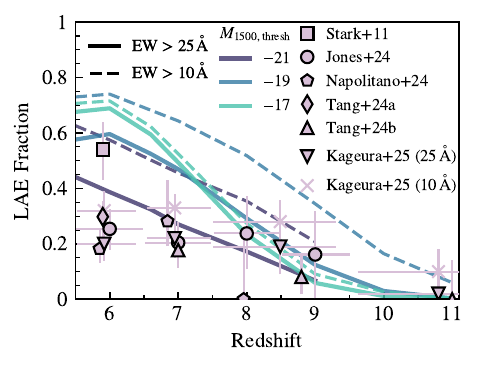}
    \caption{Fraction of Ly$\alpha$ emitters (LAEs) as a function of redshift. For the solid (dashed) curves, galaxies are considered LAEs if they have a Ly$\alpha$ equivalent width greater than 25~\AA\ (10~\AA) and an observed Ly$\alpha$ luminosity above $10^{41.5}\, \rm erg\, s^{-1}$. Each color line corresponds to a fixed upper limit in $M_{1500,\mathrm{obs}}$ such that only galaxies brighter than the threshold UV magnitude are considered. We overestimate the LAE fraction at lower redshifts compared to \textit{JWST} data (see the discussion in the text), which correspond to the EW > 25 \AA~definition of LAEs, except where noted for \cite{Kageura2025}.}
    \label{fig:fLAE_z}
\end{figure}

\begin{figure}
    \centering
    \includegraphics[width=\columnwidth]{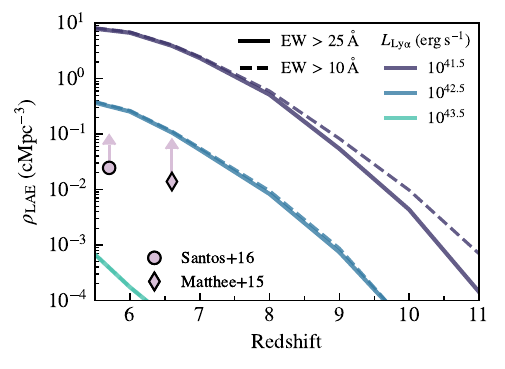}
    \caption{Density of Ly$\alpha$ emitters (LAEs) as a function of redshift. LAEs are defined as galaxies with Ly$\alpha$ equivalent widths above 25~\AA\ (10~\AA\ for dashed curves) and observed Ly$\alpha$ luminosities above the given thresholds for the colored curves. These are consistent with the observed lower limits from \citet{Santos2016, Matthee2015}.}
    \label{fig:rhoLAE_z}
\end{figure}

\subsection{Redshift evolution}
We explore how Ly$\alpha$ properties change throughout the EoR to characterize the evolution of reionization-era galaxies. Figure~\ref{fig:fescTigm_z} shows the redshift evolution of key Ly$\alpha$ quantities, \fT (as well as $f_{\rm esc}$ and $\mathcal{T}_{\rm IGM}$ separately) and EW, in galaxies brighter than certain $M_{1500}$ thresholds. The trends reveal that fainter galaxies experience a more dramatic increase in \fT as redshift decreases and the IGM becomes more ionized. In contrast, brighter galaxies tend to retain relatively lower values of \fT even at lower redshifts. This trend suggests that these galaxies are more affected by ISM processes including wind and dust attenuation throughout the EoR.

Fainter galaxies tend to exhibit higher Ly$\alpha$ EWs across all redshifts, indicating that Ly$\alpha$ emission dominates their spectra more significantly than in brighter galaxies. The overall trend of increasing EW with decreasing redshift is a direct consequence of the evolving IGM as the neutral hydrogen fraction decreases. These results underscore the role of both internal galaxy processes and the external IGM in shaping the observed Ly$\alpha$ emission properties. Fainter galaxies, while more difficult to detect, have higher EW, leading to more LAEs in those populations, especially at high redshifts. We compare to observed EW measurements from individual LAEs \citep{Matthee2018, Jung2024, Tacchella2023, Witstok2025} and find that the typical EWs in \thesan are consistent with these observations at high redshifts, with some underestimation at lower redshift, potentially due to comparing statistics to individual sources. Note that the EW at $z=10.6$ was measured by both \citet{Tacchella2023} and \citet{Bunker2023} for the same source.

Much of our understanding of the EoR comes from observations from LAEs, and the fraction of galaxies which are considered LAEs is an important statistic by which to study the evolution of reionization. We calculate the LAE fraction in \thesan by comparing the number of galaxies with observed \LLya $> 10^{41.5}\ \rm erg\, s^{-1}$ and above an EW threshold to all galaxies in the catalog. We plot the redshift evolution of the LAE fraction for EW thresholds of 25 \AA\ (solid) and 10 \AA\ (dashed) in Figure~\ref{fig:fLAE_z} considering only galaxies brighter than the UV magnitude thresholds indicated by the colors of the curves. Thus, each curve shows the fraction of bright galaxies ($M_{1500} < M_{1500,\rm thresh}$) which are LAEs to more closely correspond with biases in observations. Our predictions indicate the LAE fraction grows over time as the IGM becomes more ionized.

We compare to observed LAE fraction constraints \citep{Stark2011, Jones2024, Tang2024a, Tang2024b, Kageura2025} and find good agreement at high redshifts ($z \gtrsim 8$). Toward the end of reionization, however, we begin to overestimate the LAE fraction compared to recent \textit{JWST} observations, while matching well with the older constraint from \citet{Stark2011}. The discrepancy between our LAE fraction and the \textit{JWST} observations could be due to the early reionization history in \thesan, which could result in higher transmission at earlier redshifts than observed (see discussion of different physics variations in \thesan in Appendix~\ref{appx:thesan2}). It is also likely that there may be a methodological inconsistency between our Ly$\alpha$ continuum modeling (using BPASS and an empirical dust model) compared to the power-law continuum extrapolations performed from low and medium resolution spectroscopy or photometry \citep[see e.g. Appendix A of][]{Smith2022} The dust model may be suppressing the continuum around 1216\,\AA\ such that our calculated EWs tend to be higher than those calculated from observations, especially at later times, leading to systematically higher LAE fractions. It is not clear that empirical models are sufficiently predictive for robust Ly$\alpha$ continuum baselines, and if biases are present then directly calibrating for LAE fractions could impact other results. If we are too permissive in the LAE selection criteria this would likely be similar to adopting a different EW threshold.

In addition to LAE fraction, the spatial density of LAEs provides information about the galaxies and IGM throughout reionization. We explore the evolution of the LAE density for galaxies bright in Ly$\alpha$ (with observed Ly$\alpha$ luminosities above $10^{41.5}$ \ergs, $10^{42.5}$ \ergs, and $10^{43.5}$ \ergs) in Figure~\ref{fig:rhoLAE_z}. We compare to naive estimates of the LAE density from observational samples from \citet{Santos2016, Matthee2015}. As expected, we find that LAEs become more numerous as the EoR progresses and more of the Ly$\alpha$ signal can get through the IGM as it becomes less neutral, with very bright LAEs (\LLya$> 10^{43.5}$ \ergs) being quite rare throughout much of reionization.

\section{Connections to ionized bubble sizes}
\label{sec:bubbles}

Ionized bubble sizes are quintessential to understanding the patchy nature of reionization and its sources. With the calibrated observational properties of high-redshift \thesan galaxies, we explore connections between observables from galaxy spectra and the properties of ionized bubbles surrounding the hosts. 

We calculate the bubble sizes using the mean-free path (MFP) method, which extends rays in 768 directions from the virial radius of the central galaxy. Details of the ray tracing and bubble size studies in \thesan can be found in \citet{Neyer2024}. These lines of sight are the same for the bubble sizes and for the calculation of the IGM transmission. Thus, we incorporate the effects of different lines of sight originating from the same sources. The virial radius we use is the $R_{200,\rm crit}$ calculated for groups in the \thesan group catalogs. 

\begin{figure}
    \centering
    \includegraphics[width=\linewidth]{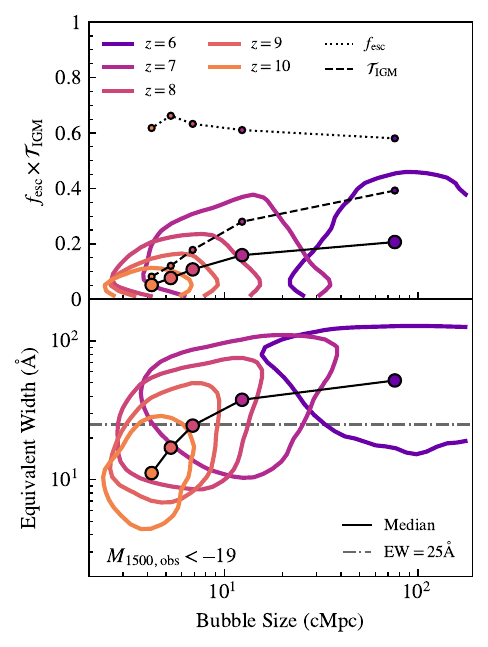}
    \caption{Upper panel: Sixty-eighth percentile contours for the product of Ly$\alpha$ escape fraction and IGM transmission (\fT) versus bubble size along each line of sight. Medians of \fT, $f_{\rm esc}$, and $\mathcal{T}_{\rm IGM}$ for each redshift are shown in colored points connected by solid, dotted, and dashed black curves, respectively. Lower panel: The same for Ly$\alpha$ equivalent width versus bubble size. All galaxies considered in both plots have observed UV magnitudes brighter than $-19$.}
    \label{fig:Rcontours}
\end{figure}

\begin{figure}  
\centering
    \includegraphics[width=\linewidth]{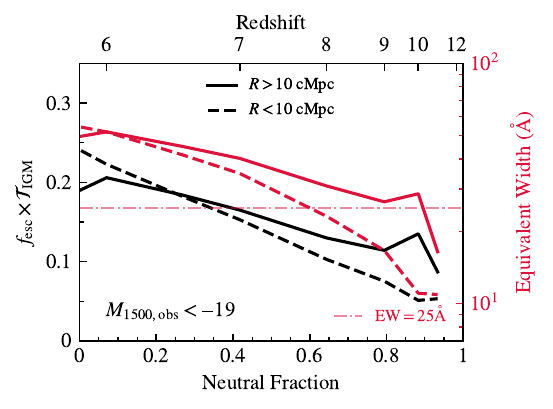}
    \caption{The product of escape fraction and IGM transmission (left axis) and Ly$\alpha$ equivalent width (right axis, in red) as a function of redshift for galaxies brighter than $M_{1500}<-19$ residing in large ($R>10$\,cMpc, solid curves) and small ($R<10$\,cMpc, dashed curves) ionized bubbles. For the brightest galaxies in the UV (or in terms of EW), the overall separation suggests a relatively modest correlation of bubble size with Ly$\alpha$ transmission and line strength. While the IGM is less than $\sim 50\%$ ionized, there tends to be stronger Ly$\alpha$ transmission coming from galaxies within large bubbles than those within smaller bubbles.}
    \label{fig:fT_EW_z_R}
\end{figure}

\begin{figure}
    \centering
    \includegraphics[width=\columnwidth]{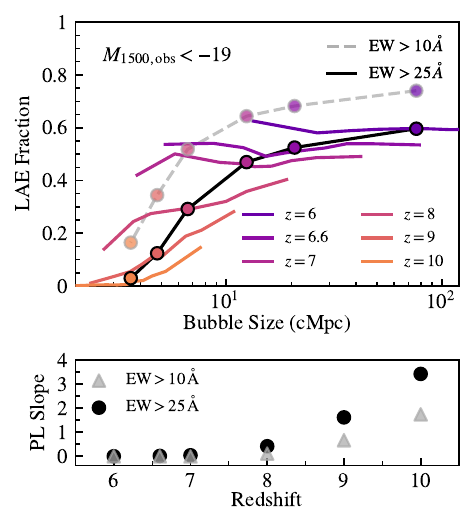}
    \caption{Upper panel: Fraction of Ly$\alpha$ emitters ($X_{\rm Ly\alpha}$) versus bubble size at various redshifts for galaxies with $M_{1500} < -19$. LAEs have $L_{\rm Ly\alpha} > 10^{41.5} \rm erg\,s^{-1}$ and EW > 10~\AA\ (dashed) or 25~\AA\ (solid). Colored points mark medians; grey and black curves connect the 10 \AA\ and 25 \AA\ cases, respectively. Lower panel: Power-law slope $\left( {\rm d}\log X_{\rm Ly\alpha} / {\rm d}\log R \right)$ versus redshift for EW > 10~\AA\ (grey triangles) and EW > 25~\AA\ (black circles), from least-squares fits to the curves above. Slopes show stronger bubble size impacts at $z \gtrsim 8$ before most bubbles in the simulation have merged.}
    \label{fig:fLAE_R}
\end{figure}

\begin{figure}
    \centering
    \includegraphics[width=\columnwidth]{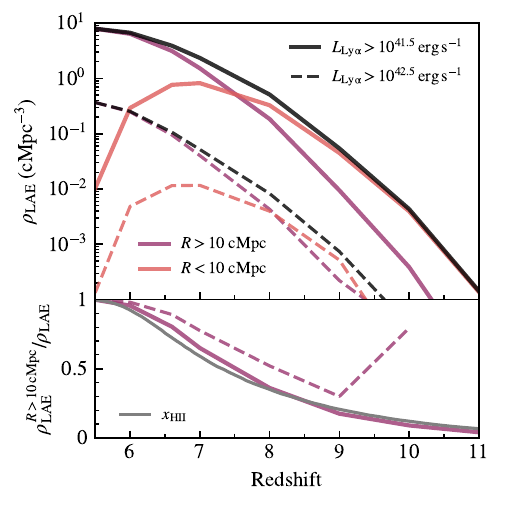}
    \caption{Upper panel: Density of LAEs in large ($R > 10\,\text{cMpc}$) and small ($R < 10\,\text{cMpc}$) ionized bubbles as a function of redshift. For the solid (dashed) curves, galaxies are considered LAEs if they have an observed Ly$\alpha$ luminosity greater than $10^{41.5}\ \rm erg\,s^{-1}$ ($10^{42.5}\ \rm erg\,s^{-1}$) and a Ly$\alpha$ equivalent width above 25 \AA. Lower panel: Ratio of $\rho_{\rm LAE}$ in large bubbles with respect to the overall $\rho_{\rm LAE}$ at the corresponding Ly$\alpha$ luminosity cut. The thin grey line shows the reionization history which closely tracks the ratio for large bubbles with $L_{\rm Ly\alpha} > 10^{41.5}\ \rm erg\,s^{-1}$.}
    \label{fig:rhoLAE_z_R}
\end{figure}

\begin{table}
\centering
\renewcommand{\arraystretch}{1.6}
\setlength{\tabcolsep}{2.7pt}
\caption{Redshift evolution of $R_{\rm bright}$, $R_{\rm faint}$, their ratio, and $W_1$ under different selection criteria.}
\begin{tabular}{p{1.82cm}@{\hskip -10pt}lcccc}
\hline
\textbf{Quantity} & \textbf{Criteria} & $z=6$ & $z=7$ & $z=8$ & $z=10$ \\
\hline

\multirow{3}{1.82cm}{$R_{\rm bright}$\\[1pt] (cMpc)}
 & $M_{1500} < -19$                & $76.1^{+108.9}_{-48.5}$ & $12.4^{+12.7}_{-6.5}$ & $6.6^{+5.4}_{-2.8}$ & $3.6^{+1.5}_{-1.1}$ \\
 & $L_{\rm Ly\alpha} > 10^{42}$    & $76.7^{+109.1}_{-49.0}$ & $14.5^{+14.2}_{-7.6}$ & $8.6^{+6.6}_{-3.7}$ & $5.6^{+3.1}_{-1.6}$ \\
 & $\mathrm{EW} > 25$\,\AA\        & $72.3^{+109.4}_{-49.3}$ & $10.9^{+13.5}_{-6.7}$ & $6.7^{+6.1}_{-3.2}$ & $5.7^{+2.6}_{-2.2}$ \\
\hline

\multirow{3}{1.82cm}{$R_{\rm faint}$\\[1pt] (cMpc)}
 & $M_{1500} > -19$                & $72.2^{+109.4}_{-49.2}$ & $10.3^{+13.4}_{-6.6}$ & $4.8^{+5.6}_{-2.6}$ & $2.0^{+1.7}_{-0.9}$ \\
 & $L_{\rm Ly\alpha} < 10^{42}$    & $71.9^{+109.4}_{-49.1}$ & $10.1^{+13.2}_{-6.5}$ & $4.8^{+5.6}_{-2.6}$ & $2.1^{+1.7}_{-0.9}$ \\
 & $\mathrm{EW} < 25$\,\AA\        & $73.3^{+109.1}_{-48.7}$ & $9.3^{+12.8}_{-6.2}$ & $4.0^{+5.1}_{-2.1}$ & $2.0^{+1.6}_{-0.9}$ \\
\hline

\multirow{3}{1.82cm}{$\dfrac{R_{\rm bright}}{R_{\rm faint}}$}
 & $M_{1500}$                      & $1.05$ & $1.21$ & $1.39$ & $1.77$ \\
 & \LLya              & $1.07$ & $1.43$ & $1.8$ & $2.72$ \\
 & $\mathrm{EW}$                   & $0.99$ & $1.18$ & $1.66$ & $2.79$ \\
\hline

\multirow{3}{1.82cm}{$W_1(\log R)$\\[1pt] (dex)}
 & $M_{1500}$                      & $0.05$ & $0.11$ & $0.15$ & $0.24$ \\
 & \LLya              & $0.05$ & $0.18$ & $0.26$ & $0.44$ \\
 & $\mathrm{EW}$                   & $0.01$ & $0.08$ & $0.21$ & $0.42$ \\
\hline
\label{tab:R_gal}
\end{tabular}
\end{table}

\begin{figure*}
    \centering
    \includegraphics[width=\linewidth]{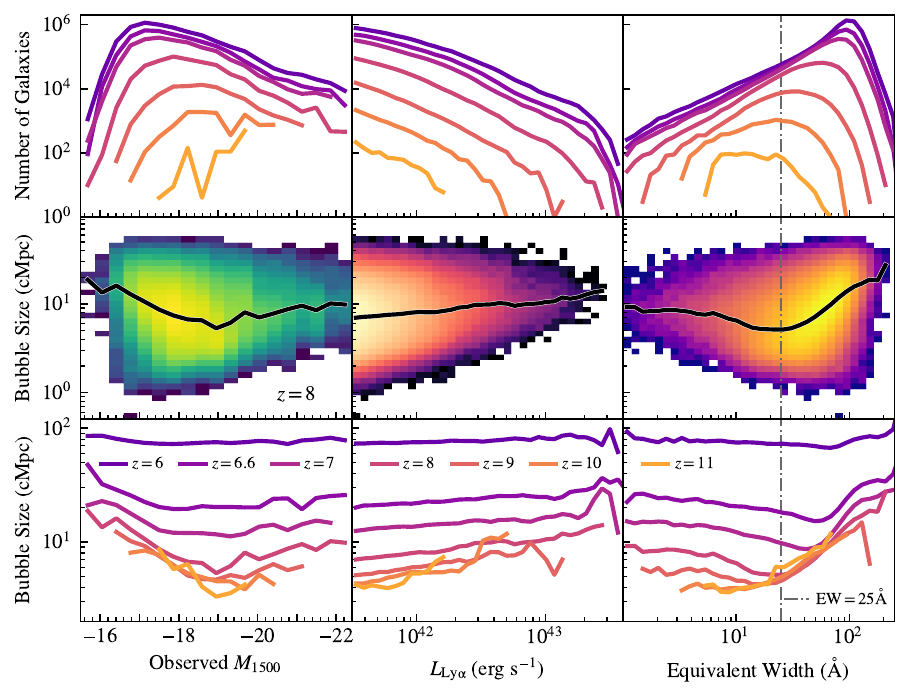}
    \caption{Ionized bubble size versus observed UV magnitude (left column), observed Ly$\alpha$ luminosity (center column), and equivalent width (right column) to examine the relationship between bubble sizes and observability. The upper row shows the number of galaxies in our sample with observed \LLya > $10^{41.5}$ \ergs. The center row shows two-dimensional histograms for $z=8$, and solid curves trace median bubble sizes for several redshifts in the lower panels. There is a general trend of flattening of the medians as the EoR progresses indicating that early trends with UV magnitude, Ly$\alpha$ luminosity, and EW largely wash out by the end of reionization. EW has the strongest correlation with bubble size, persisting even down to $z = 6.6$.}
    \label{fig:2d_R_MLEW}
\end{figure*}

\subsection{Impact of bubble sizes on Ly$\alpha$ properties}
\label{subsec:lya_bubbles}

We study the relationship between observed Ly$\alpha$ properties of galaxies and the size of the ionized bubble along the line of sight. The distributions of \fT and EW with respect to bubble size for galaxies with observed UV magnitudes brighter than $-19$ are characterized by 1-$\sigma$ contours at a variety of redshifts in Figure~\ref{fig:Rcontours}. The overall medians for each redshift are shown in the colored circles and connected to indicate the overall redshift trends as bubbles grow and observed Ly$\alpha$ emission becomes stronger. In the upper panel, we also show the evolution of the medians for just $f_{\rm esc}$ and $\mathcal{T}_{\rm IGM}$ connected by dotted and dashed curves, respectively. The distributions broaden and move to higher bubble sizes, \fT, and EW as more gas is ionized allowing more Ly$\alpha$ transmission. We see a pile up of galaxies near the maximum bubble size at redshift 6 which also has lower \fT and EW. These are likely contributions from the brightest galaxies which are within large bubbles by $z=6$, and have lots of dust which suppresses the Ly$\alpha$ transmission.

We explore further the redshift evolution with Figure~\ref{fig:fT_EW_z_R} which plots the median values of \fT and EW for galaxies within large ionized bubbles ($R>10\,\rm cMpc$, solid curves) and small ionized bubbles ($R<10\,\rm cMpc$, dashed curves). There is an overall trend of increasing \fT and EW as reionization progresses regardless of bubble size. At times before the midpoint of reionization ($x_{\rm HI} > 0.5$, $z\gtrsim 7$), there is stronger Ly$\alpha$ transmission from the galaxies within larger bubbles, indicated by the higher \fT and EW. Early ionized bubbles form around bright galaxies and expand such that Ly$\alpha$ photons can redshift out of resonance before encountering the neutral IGM. Thus, a large bubble size is necessary at early times to allow high IGM transmission and equivalent widths. However, as bubbles merge and the IGM becomes increasingly ionized, a larger variation of galaxies, including those with low escape fractions due to dust and other factors become encompassed in large bubbles, leading to a relative decrease in \fT and EW compared to their counterparts in smaller bubbles.

\begin{figure*}
    \centering
    \includegraphics[width=\linewidth]{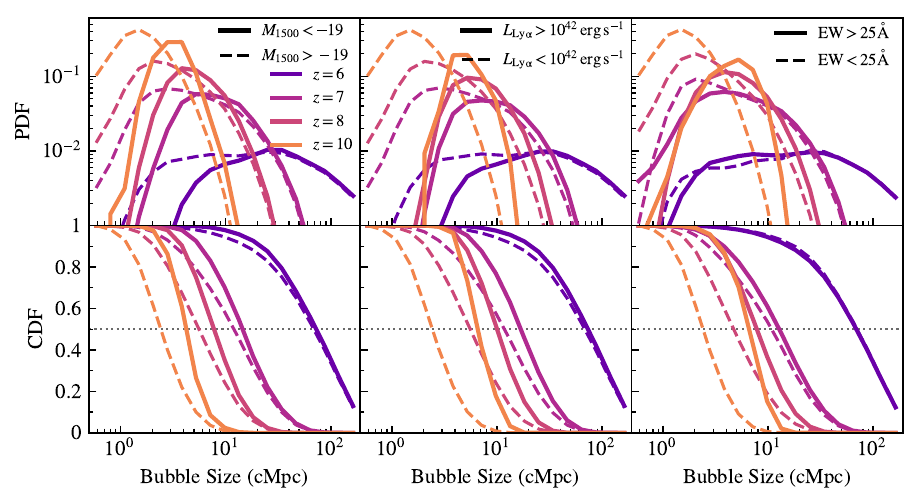}
    \caption{Distributions of bubble sizes around bright (solid curves) and faint (dashed curves) galaxies according to three different criteria: $M_{1500}$, \LLya, and EW. The upper row plots the probability density function (PDF $= \text{d}P/\text{d}R$) and the lower row shows the cumulative distribution function (CDF $=P(>R)$).The bubbles around bright galaxies are preferentially larger than those around fainter galaxies for all three criteria.}
    \label{fig:LAE_R_dists}
\end{figure*}

We expect ionized bubbles to impact LAE visibility by allowing more peak Ly$\alpha$ photons to redshift out of resonance before encountering neutral gas, resulting in higher observed EWs. We investigate the relationship between LAE fraction and bubble sizes throughout the EoR in Figure~\ref{fig:fLAE_R}. The upper panel shows the LAE fraction versus bubble size in colored curves, with the colored points indicating the overall median LAE fraction and bubble size at that redshift. The black line connects these medians to emphasize the trend of increasing bubble sizes and increasing LAE fraction at reionization takes place. The fainter points and dashed curve indicate the trend in medians for LAEs with EW $> 10$ \AA, while the solid curves require EW$> 25$ \AA. We characterize the correlation between LAE fraction and bubble size at each redshift by performing a least-squares power law fit and plotting the resulting slope $\left( {\rm d}\log X_{\rm Ly\alpha} / {\rm d}\log R \right)$ in the lower panel of Figure~\ref{fig:fLAE_R}. As reionization progresses, the power law slopes decrease indicating that the correlation between LAE fraction and bubble sizes becomes less pronounced. At early times, there are proportionally more LAEs in larger bubbles since very small bubbles don't allow photons to shift out of resonance before being absorbed by surrounding neutral gas.

We are also interested in how the density of LAEs changes for galaxies in differently sized bubbles. We compare the LAE density throughout the EoR for galaxies in large ($R > 10$ cMpc, purple curves) and small ($R < 10$ cMpc, pink curves) bubbles to the overall evolution of the LAE density in the \thesan box (black curves) in the upper panel of Figure~\ref{fig:rhoLAE_z_R}. At early redshifts when most bubbles are below 10 cMpc in size, there are more LAEs in these small bubbles. However, around $z \sim 7.5$, there is a transition to a higher density of LAEs residing in larger bubbles. We also examine the impact of observed Ly$\alpha$ luminosity on these results, seeing a qualitatively similar trend with lower overall densities for brighter LAEs (\LLya $> 10^{42.5}$ \ergs, dashed curves). In the lower panel of Figure~\ref{fig:rhoLAE_z_R}, the ratio of LAE density in large bubbles to overall LAE density is compared to the ionization history of \thesan (grey curve). This ratio closely tracks the reionization history when accounting for all LAEs with \LLya $> 10^{41.5}$ \ergs. The upturn in the ratio for \LLya $> 10^{42.5}$ \ergs at $z \gtrsim 9$ is likely due to early bright LAEs necessarily residing within large bubbles in order to have such a high observed \LLya when the IGM is largely neutral.

\begin{figure*}
    \centering
    \includegraphics[width=\linewidth]{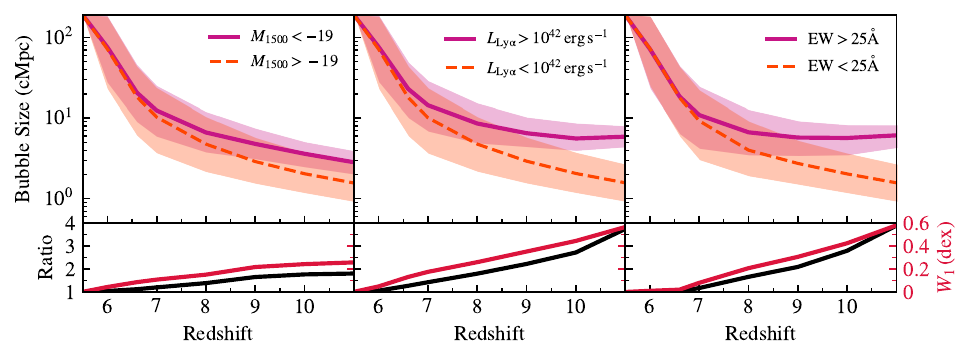}
    \caption{Ionized bubble size versus redshift for bright galaxies (magenta solid curves) and faint galaxies (orange dashed curves), according to an $M_{1500}$ cut (left), Ly$\alpha$ luminosity cut (center), and equivalent width cut (right). The bottom panels show the ratio of $R_{\rm bright}/R_{\rm faint}$ in black and the Wasserstein distance between the bright and faint logarithmic bubble size distributions, $W_1(\log R)$ in red. Galaxies that are bright in Ly$\alpha$ luminosity and those that have high Ly$\alpha$ EWs are more likely to be in larger bubbles than their fainter counterparts, especially at higher redshifts.}
    \label{fig:R_z_bright}
\end{figure*}

\subsection{Observational probes of bubble sizes}
\label{subsec:probes}

Currently, the primary way to constrain ionized bubble sizes observationally is through careful analysis of high-$z$ galaxy properties. The observed spectra from quasars are also incredibly effective as probes of reionization, but the relative rarity of these objects reduces the practicality of studying these constraints in the \thesan box. To this end, we investigate the relationships between high-$z$ galaxies and bubble sizes through three key galaxy observables: UV magnitude, Ly$\alpha$ luminosity, and Ly$\alpha$ equivalent width. In the top panels of Figure~\ref{fig:2d_R_MLEW}, we show the number of galaxies in \thesan with observed \LLya > $10^{41.5}$ \ergs according to each of these properties. By making this \LLya cut, the sampled galaxies are biased to higher UV magnitudes, particularly earlier in reionization, and to higher EWs. To illustrate the connection between bubble sizes and the three galaxy parameters ($M_{1500}$, \LLya, and EW), two-dimensional distributions are shown for $z=8$ in the center row of Figure~\ref{fig:2d_R_MLEW}, with the black curves showing the median bubble sizes. The lower panels show the medians for several redshifts as the relationships evolve throughout the EoR. In all cases, the correlation weakens as reionization occurs with the median bubble sizes being nearly constant for all galactic properties by $z \lesssim 6$. Earlier in reionization, however, there are clear trends in the Ly$\alpha$ properties indicating that galaxies with higher \LLya or EW reside in preferentially larger bubbles, especially at particularly high EW (EW $\gtrsim 25$ \AA). This trend also appears to a lesser extent for UV-bright galaxies ($M_{1500} \lesssim -19$) which are more highly sampled under the \LLya conditions.

To further explore how bubbles around galaxies differ based on these galactic properties, we divide the \thesan galaxies into bright and faint groups according to criteria based on the three properties divided at ($M_{1500} = -19$, \LLya $= 10^{42}$ \ergs, EW $= 25$ \AA). We report the median bubble sizes and 16$^\text{th}$ to 84$^\text{th}$ percentiles for the bright and faint groups, as well as their ratio, and the Wasserstein distance, which quantifies the difference between the logarithmic bubble size distributions, in Table~\ref{tab:R_gal}. The distributions for bubbles around bright (solid curves) and faint (dashed curves) galaxies are shown in Figure~\ref{fig:LAE_R_dists} for the three criteria at a few key redshifts, with the probability density function (PDF $= \text{d}P/\text{d}R$) in the upper panels and the cumulative distribution function (CDF $=P(>R)$) in the lower panels. In general, the distributions for the bright galaxies are shifted to higher bubble sizes than for faint galaxies, with the difference at each redshift lessening as reionization progresses. The largest difference between the bubbles sizes around bright and faint galaxies occurs with the \LLya criteria. While EW starts as an effective discriminator for bubble size populations, the distributions rapidly become quite similar throughout reionization.

The redshift evolution of the characteristic bubble sizes around bright and faint galaxies is explored in Figure~\ref{fig:R_z_bright} with the median bubble sizes around bright (solid magenta) and faint galaxies (dashed orange), as well as correspondingly colored 16$^\text{th}$ to 84$^\text{th}$ percentile shaded regions for each of the criteria in the upper panels. The lower panels show the ratio ($R_{\rm bright} / R_{\rm faint}$, black) and the Wasserstein distance between the distributions of $\log R$ ($W_1$, red). In all cases, the bubbles around bright galaxies are preferentially larger than those around faint galaxies, with the difference in bubble size populations decreasing as reionization progresses. Both the \LLya and EW criteria correspond with two distinct populations of bubble sizes at higher redshifts as bubbles are just starting to form and grow. By $z \lesssim 7$, the bubble size populations around high and low EW galaxies have converged to nearly identical, while the characteristic bubble sizes around high and low \LLya galaxies, and to a lesser extent around UV-bright and faint galaxies, have remained slightly more distinct. With this, we conclude that \LLya is the most illustrative galaxy property of $M_{1500}$, \LLya, and EW in determining typical bubble sizes surrounding the galaxies.

\section{Conclusions}
\label{sec:conclusions}

The properties and visibility of high-redshift galaxies are critical for informing our understanding of the EoR. Ionized bubbles, while not currently directly observable, encode important information about the progression of reionization through the IGM as well as the sources which drive it. In turn, the sources themselves allow for a window into constraining ionized bubble sizes through analysis of their properties and spectra, specifically through observations of Ly$\alpha$ emission from the galaxies. In this paper, we used the cosmological radiation-hydrodynamic simulation suite, \thesan to better understand the connections between galactic Ly$\alpha$ properties and the ionized bubble in which the galaxies reside. We use the intrinsic galaxy Ly$\alpha$ luminosities and IGM transmission calculated directly from \thesan and implement a simple model to account for unresolved galactic-scale physics including dust, wind, and velocity offsets in order to calibrate Ly$\alpha$ escape fractions and observed Ly$\alpha$ luminosities for galaxies in \thesan. Using these calibrated quantities, we explore properties of LAEs and connect them to the ionized bubbles surrounding them.

We describe the main points of this work below:
\begin{enumerate}[itemsep=1pt, topsep=2pt, parsep=1pt]
    \item We present an improved calibrated model for Ly$\alpha$ properties of galaxies in \thesan from our previous study in \citet{Xu2023}. The trends are generally consistent with observations and allow for more robust comparisons between the simulation and observed measurements and constraints.
    \item The fraction of UV-detected galaxies that are LAEs (LAE fraction) predicted from the calibrated galaxy properties is consistent with \emph{JWST} measurements early in reionization ($z \gtrsim 7$), while overpredicting the LAE fraction at lower redshifts as reionization concludes slightly too early in \thesan.
    \item The LAE fraction corresponds with bubble size before the midpoint of reionization ($z \gtrsim 7$) as LAE visibility depends strongly on the ionized gas immediately surrounding the galaxy as well as the global IGM conditions.
    \item There is a strong correlation with bubble size and LAE properties, e.g. Ly$\alpha$ luminosity and Ly$\alpha$ equivalent width, particularly during the first half of reionization. However, this is washed out once bubbles overlap such that faint and bright, allowing fainter galaxies to become LAEs at lower redshifts.
\end{enumerate}

In summary, we have developed an updated calibration model for Ly$\alpha$ observables of \thesan galaxies and studied implications for the connections between galaxy properties and ionized bubble sizes. Observational bubble size constraints are primarily measured through analysis of Ly$\alpha$ emission from reionization-era galaxies as the details of the transmission encodes information about the local ionized region. By examining the properties of bright galaxies in \thesan, particularly LAEs, in relation to the ionized bubbles surrounding them, this work begins to bridge the gap between theoretical bubble size predictions and observable feasibility. 

Further work could explore the particular effect of bubble sizes on the Ly$\alpha$ damping wing, which is one of the most prominent observational indications of bubble size. Synthetic rest-frame Ly$\alpha$ spectra from \thesan galaxies could provide more comparable bubble size estimates to current observed constraints. Additionally, examining clustering of LAEs with respect to the underlying densities and bubbles would provide another window into the processes driving reionization. The catalog of calibrated galaxy properties can be used for a variety of additional science cases to improve observability of theoretical reionization predictions from \thesan. 

\section*{Acknowledgments}
We thank the anonymous referee for their helpful comments and suggestions. AS acknowledges support through HST AR-17859, HST AR-17559, and JWST AR-08709. MV acknowledges support through NSF AAG AST-2408412, JWST AR-04814, NASA ATP 21-ATP21-0013, NASA ATP 21-ATP21-0013 and NSF AAG AST-2307699. LK acknowledges the support of a Royal Society University Research Fellowship (grant number URF$\backslash$R1$\backslash$251793).

\section*{Data Availability}

Data products from \thesan are available online at \href{https://www.thesan-project.com}{www.thesan-project.com} as described in \citet{Garaldi2023}. The updated Ly$\alpha$ galaxy catalogs will be made available on the \thesan website upon publication.



\bibliographystyle{mnras}
\bibliography{biblio} 




\appendix
\section{Alternative physics variations}
\label{appx:thesan2}
\begin{figure*}
    \centering
    \includegraphics[scale=1]{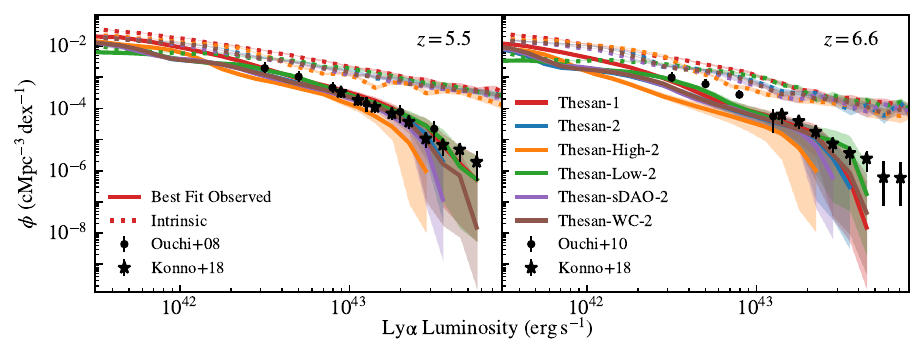}
    \caption{Ly$\alpha$ luminosity function at $z=5.5$ (left panel) and $z=6.6$ (right panel) using the \thesanone best-fit model applied to different \thesan physics models. The model with low-mass haloes contributing to reionization (green curve) provides a good match to observations across both redshifts, whereas the model with high-mass haloes contributing (orange curve) consistently underpredicts the number of bright galaxies.}
    \label{fig:bestfit_thesan2}
\end{figure*}

\begin{figure}
    \centering
    \includegraphics[width=\linewidth]{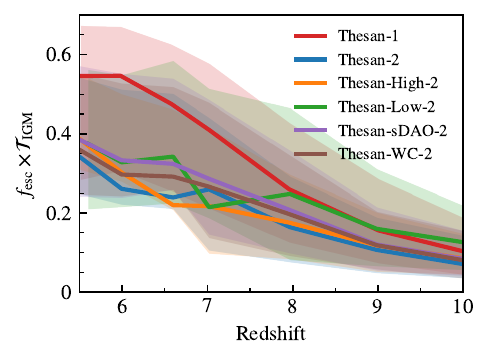}
    \caption{Escape fraction times IGM transmission versus redshift for each of the \thesan runs. We only consider contributions from lines of sight with observed Ly$\alpha$ luminosity $> 10^{41.5}\,\rm erg\, s^{-1}$. Median \fT values are lower in the \thesantwo variants than in \thesanone due to poorly-resolved faint galaxies in the lower resolution \thesantwo simulations.}
    \label{fig:fescTigm_xHI_T2}
\end{figure}

\begin{figure*}
    \centering
    \includegraphics[width=\linewidth]{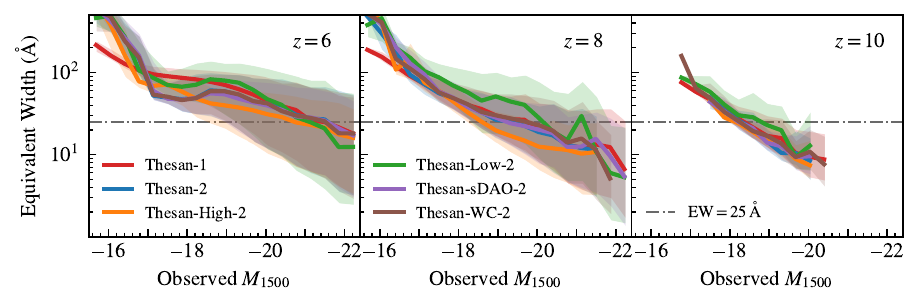}
    \caption{Ly$\alpha$ equivalent width versus observed UV magnitude. The dashed line shows the common EW threshold of 25\,\r{A}. Different redshifts are shown in different panels and the colored curves correspond to different \thesan simulations. Across all simulations, the EWs are consistent with UV-bright galaxies having lower Ly$\alpha$ EWs.}
    \label{fig:EW_T2}    
\end{figure*}

Here we explore the impact of the updated calibration parameters on Ly$\alpha$ properties in the \thesantwo runs with different physical processes, including different reionization sources and alternative dark matter. These are using the same best-fit model as before (i.e. the one fit using \thesanone st $z=5.5$ and $z=6.6$ simultaneously). All \thesantwo simulations are eight times lower resolution than \thesanone, with \thesantwo using the same physics modeling as \thesanone, and \thesanwc imposing weak numerical convergence on the neutral fraction as a function of redshift. \thesanhigh only considers contributions to reionization from high mass haloes $\left( >10^{10} \right) \Msun$, and \thesanlow only considers contributions from low mass haloes $\left( <10^{10} \right) \Msun$. \thesansdao implements an alternative dark matter model with strong dark acoustic oscillations.

Figure~\ref{fig:bestfit_thesan2} shows the Ly$\alpha$ LFs for $z=5.5$ (left panel) and $z=6.6$ (right panel) for multiple \thesan runs, derived from the best-fit parameters from \thesanone (compiled in Table~\ref{tab:bestfit}). \thesanlow, which only considers low mass haloes for reionization, aligns particularly well with observed data at both redshifts supporting the idea that faint, low-mass galaxies play a crucial role in driving reionization, contributing significantly to the overall Ly$\alpha$ luminosity output. In contrast, \thesanhigh, which assumes high-mass haloes as the primary drivers of reionization, tends to underpredict the LF. This discrepancy suggests that high-mass galaxies alone cannot account for the full population of Ly$\alpha$ emitting galaxies. The contribution from lower-mass systems appears essential for reproducing the observed abundance of Ly$\alpha$ emitting galaxies, underscoring the importance of these small galaxies during the reionization epoch.

At the bright end of the LFs, we observe a pronounced drop-off due to the limited volume of the \thesan simulation box, which reduces the statistical likelihood of capturing rare, highly luminous galaxies. Despite this limitation, the simultaneous fit across both redshifts highlights the robustness of the model with low-mass halo contributions in matching the observations. The bright-end behavior, however, remains underpredicted, likely due to volume constraints rather than physical discrepancies, in both \thesanone and \thesantwo simulations.

We explore the evolution of \fT\ over time for the different physics variations in Figure~\ref{fig:fescTigm_xHI_T2}. As expected, the proportion of Ly$\alpha$ photons that are able to escape the galaxies and travel through the IGM to be observed increases as reionization progresses and there is less neutral gas to absorb the radiation. While all the \thesantwo simulations have similar values of \fT to each other, they are significantly lower than those in \thesanone for $z \lesssim 8$. This discrepancy is likely due to the impact of small, faint galaxies with high values of \fT, which are resolved in \thesanone, but not in \thesantwo. If we include only Ly$\alpha$-bright galaxies (e.g. \LLya $> 10^{42}$ \ergs), then the \thesanone curve more closely matches the \thesantwo curves.

The typical Ly$\alpha$ equivalent widths for different UV magnitudes at different redshifts are shown in Figure~\ref{fig:EW_T2}. The typical EW observed from a galaxy is lower for galaxies brighter in the UV. This trend is robust to physics and resolution changes, with only small deviations. Most notable of these slight deviations are the generally higher EWs in \thesanlow indicating that the low-mass galaxies yield relatively stronger Ly$\alpha$ emission and higher EWs. Conversely, \thesanhigh appears to have slightly lower typical EWs than the other runs, indicating that more massive galaxies have lower EW, likely due to higher  attenuation from dust and winds.

\appendix
\section{3D Sizes of Bubbles}
\label{appx:3dbubbles}
\begin{figure}
    \centering
    \includegraphics[width=\linewidth]{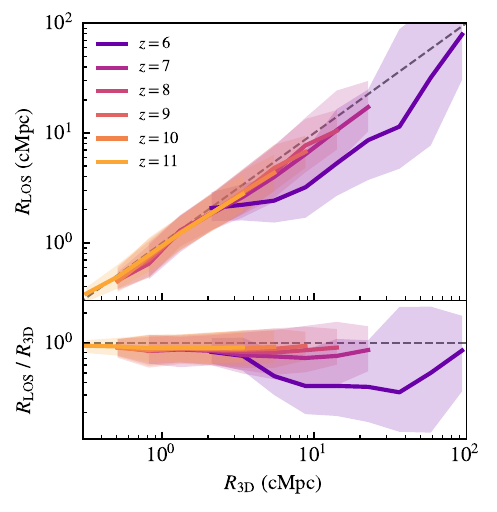}
    \caption{Comparison between bubble sizes along lines of sight ($R_{\rm LOS}$) versus the average 3D bubble radius from the central galaxy's simulation cell ($R_{\rm 3D}$). The upper panel shows the median and 16$^\text{th}$ to 84$^\text{th}$ percentile ranges of $R_{\rm LOS}$ from galaxies within an ionized bubble of average radius $R_{\rm 3D}$. The lower panel shows the ratio for more direct comparison. Naturally, there is some spread due to sightline-to-sightline variations, but in general, $R_{\rm LOS}$ traces the $R_{\rm 3D}$ well, especially for $z \gtrsim 7$.}
    \label{fig:R3d}
\end{figure}

In this work, we consider bubble sizes along particular lines of sight from galaxies, corresponding with sightline-dependent properties including IGM transmission, calibrated Lyman-$\alpha$ luminosity, and equivalent width. However, due to the complex morphology of ionized bubbles, these line-of-sight bubble sizes ($R_{\rm LOS}$) do not explicitly encapsulate the 3D sizes of the ionized bubbles. Thus, in this appendix, we investigate the relationship between the bubble size along a line of sight and the average 3D bubble radius ($R_{\rm 3D}$).

We calculate the 3D bubble size using the mean-free path method described in \citet{Neyer2024}. For each ionized simulation cell, we perform second-order ray tracing along 192 isotropically distributed lines of sight to find the distance to the nearest neutral ($x_{\rm HII} < 0.5$) cell in that direction. We then take the average of these 192 ray lengths to be $R_{\rm 3D}$ for that simulation cell.

The line-of-sight bubble sizes are calculated in a similar manner, with the ray tracing starting one virial radius away from each galaxy center. The virial radii are calculated to be the distance from the center of the dark matter halo (as identified by the friends-of-friends halo finder algorithm) such that mean density is 200 times the critical density of the Universe. We calculate the bubble size ($R_{\rm LOS}$) along each of the 768 sightlines from each galaxy in the catalog, corresponding with the lines of sight along which the Ly$\alpha$ properties are calculated.

The comparison between $R_{\rm LOS}$ and $R_{\rm 3D}$ is shown in Figure~\ref{fig:R3d}. The upper panel shows the median and 16$^\text{th}$ to 84$^\text{th}$ percentile ranges for the values $R_{\rm LOS}$ from galaxies within a simulation cell with average bubble size $R_{\rm 3D}$. The lower panel shows the ratio of $R_{\rm LOS}$ to $R_{\rm 3D}$. As expected, there are variations among sightlines, with $R_{\rm LOS}$ tracing $R_{\rm 3D}$ well, especially at $z \gtrsim 7$. The spread can be explained by the irregular morphology of ionized bubbles and location of galaxies within ionized bubbles. We note that the $R_{\rm LOS}$ median is systematically lower than $R_{\rm 3D}$ because $R_{\rm 3D}$ is an average which relatively over-weights large bubble sizes, while the $R_{\rm LOS}$ median is agnostic to extreme bubble sizes. Thus, the more pronounced discrepancy between the median of $R_{\rm LOS}$ compared to $R_{\rm 3D}$ at $z \lesssim 7$ can be attributed to the median accounting for many short ray lengths while $R_{\rm 3D}$ is an average which is skewed larger by the long rays through the largest bubbles. The averages of $R_{\rm LOS}$ generally match $R_{\rm 3D}$ to within about 5 percent. Thus, while the average $R_{\rm LOS}$ remains a robust tracer of the average 3D bubble size, the breakdown in agreement between the median $R_{\rm LOS}$ and $R_{\rm 3D}$ at late times ($z \sim 6$) indicates that typical line-of-sight measurements should be interpreted with care late in reionization, particularly since this is where observational data are currently best positioned to infer such measurements.


\end{document}